\documentclass[twocolumn]{revtex4}

\usepackage{graphicx}
\usepackage{bm}
\usepackage{subfigure}
\usepackage{mathtools}
\usepackage{color}
\usepackage{paralist}
\usepackage{amssymb,amsfonts,amsmath}%
\usepackage[dvipsnames,usenames]{xcolor}%
\usepackage{array,booktabs,units}
\usepackage[mathscr]{euscript}
\usepackage{cancel}

\graphicspath{{./fig/}{./png/}}

\newcommand{\alphaem}{\ensuremath{\alpha_{\rm em}}}

\def\kB{k_{\rm B}}

\newcommand{\nab}{\bm{\nabla}}
\newcommand{\AAA}{\bm{A}}
\newcommand{\BB}{\bm{B}}

\newcommand{\EE}{\bm{E}}
\newcommand{\JJ}{\bm{J}}
\newcommand{\UU}{\bm{U}}

\newcommand{\SSSS}{\mbox{\boldmath ${\sf S}$} {}}
\newcommand{\meanrho}{\overline{\rho}}

{}
{}
{}

{}
{}
{}
{}
{}
{}
{}
{}
{}
{}
{}
{}
{}
{}
{}
{}
{}

{}
{}
{}

\newcommand{\meanAB}{\langle\mathcal{H}\rangle}
\newcommand{\EQA}{\begin{eqnarray}}
\newcommand{\ENA}{\end{eqnarray}}

\newcommand{\bra}[1]{\langle #1\rangle}

\def\Pm{{\rm Pr}_{_\mathrm{M}}}
\def\Rm{{\rm Re}_{_\mathrm{M}}}
\def\Rey{{\rm Re}_{_\mathrm{K}}}
\def\RmABmin{{\rm Re}_{_\mathrm{M}}^{{\mathrm{min}(\mathcal{H})}}}
\def\Rmkpkone{{\rm Re}_{_\mathrm{M}}^{{k_\mathrm{p}=1}}}
\def\Rmmax{{\rm Re}_{_\mathrm{M}}^{\mathrm{max}}}

\def\kp{k_\mathrm{p}}

\def\cs{c_{\rm s}}

\usepackage{hyperref}
\newcommand\myshade{85}
\colorlet{mylinkcolor}{violet}
\colorlet{mycitecolor}{Aquamarine}
\colorlet{myurlcolor}{YellowOrange}

\usepackage{ulem}

\def\dd{\mathrm{d}}

\hypersetup{
 linkcolor  = mylinkcolor,
 citecolor  = mycitecolor!\myshade!black,
 urlcolor   = myurlcolor!\myshade!black,
colorlinks = true,
}

\begin{document}

\title{Generation of chiral asymmetry via helical magnetic fields}

\author{Jennifer Schober}
\email{jennifer.schober@epfl.ch}
\affiliation{Laboratoire d'Astrophysique, EPFL, CH-1290 Sauverny, Switzerland}

\author{Tomohiro Fujita}
\email{t.fujita@tap.scphys.kyoto-u.ac.jp}
\affiliation{Department of Physics, Kyoto University, Kyoto, 606-8502, Japan}

\author{Ruth Durrer}
\email{ruth.durrer@unige.ch}
\affiliation{D\'epartment de Physique Th\'eorique and Center for Astroparticle Physics,
Universit\'e de Gen\`eve, Quai E. Ansermet 24, CH-1211 Gen\`eve 4, Switzerland}

\medskip
\medskip

%\submitted{\today}
\begin{abstract}
%Background
It is well known that helical magnetic fields undergo a so-called inverse
cascade by which their correlation length grows due to the conservation of
magnetic helicity in classical ideal magnetohydrodynamics (MHD). 
At high energies above approximately $10$ MeV, however, classical MHD is necessarily 
extended to chiral MHD and then the conserved quantity is 
$\meanAB + 2 \langle\mu_5\rangle / \lambda$ with $\meanAB$ being the mean 
magnetic helicity and $\langle\mu_5\rangle$ being the mean chiral chemical potential of charged fermions.
Here, 
$\lambda$ is a (phenomenological) chiral feedback parameter.
%Research Target
In this paper, we study the evolution of the chiral MHD system with the initial 
condition of nonzero $\meanAB$ and vanishing $\mu_5$. 
We present analytic derivations for the time evolution of $\meanAB$ 
and $\langle\mu_5\rangle$ that we compare
to a series of laminar and turbulent three-dimensional direct numerical simulations.
%Main Result
We find that the late-time evolution of $\meanAB$ depends on the magnetic and kinetic
Reynolds numbers $\Rm$ and $\Rey$. 
For a high $\Rm$ and $\Rey$ where turbulence occurs, $\meanAB$
eventually evolves in the same way as in classical ideal MHD where
 the inverse correlation length of the helical magnetic field scales with time $t$ as 
$k_\mathrm{p} \propto t^{-2/3}$.
For a low Reynolds numbers where the velocity field is negligible, the scaling is changed 
to $k_\mathrm{p} \propto t^{-1/2}\mathrm{ln}\left(t/t_\mathrm{log}\right)$.
After being rapidly generated, $\langle\mu_5\rangle$ always decays together with
$k_\mathrm{p}$, i.e.\ $\langle\mu_5\rangle \approx k_\mathrm{p}$, with a time evolution that depends on whether the 
system is in the limit of
low or high Reynolds numbers.
\end{abstract}


\maketitle

%%%%%%%%%%%%%
% Section 1 %
%%%%%%%%%%%%%

\section{Introduction}
\label{sec_intro}

% MHD in nature
Natural systems can be modeled as fluids when their macroscopic spatial 
extension is much larger than the typical mean free path of particle collisions.
This often applies in geophysics and astrophysics including planets, 
stars, the interstellar medium,
and galaxies. 
Hydrodynamics is even applicable in cosmology, in particular, when modeling the plasma of the early
Universe and the early stages of cosmic structure formation. 
Often, natural fluids are highly turbulent which is quantified by 
large hydrodynamic Reynolds numbers $\Rey$.
The latter measures the ratio of advection and diffusion effects and is defined as
$\Rey \equiv u_\mathrm{rms}/(k_\mathrm{f} \nu)$ where $u_\mathrm{rms}$ is the rms
velocity, $k_\mathrm{f}$ is the forcing wave number, and $\nu$ is the viscosity. 
Indeed, $\Rey \gg 1$ in many astrophysical fluids 
resulting from an efficient turbulent driving
that has its origin, for instance in galaxies, in supernova explosions 
\citep{MacLowKlessen2004}
and/or accretion flows \citep{KlessenHennebelle2010}.
Moreover, the local Universe appears to be permeated with magnetic fields
on all length scales accessible to observations:
They span from planets \cite{Stevenson2003} and 
stars \cite{DonatiLandstreet2009}, including our
Sun \cite{SolankiEtAl2006}, to the interstellar medium \cite{Crutcher2012}, 
galaxies \cite{Beck2012}, up to galaxy clusters \cite{GovoniFeretti2004} 
and possibly cosmic voids~\cite{Neronov:1900zz}.
% These astrophysical magnetic fields can be strong enough to
% be dynamical important.
The most established mechanisms to amplify and maintain magnetic
fields in stars and galaxies are magnetohydrodynamical (MHD) 
dynamos that convert kinetic energy into magnetic energy \citep{BS05}.
In general, MHD turbulence describes the dynamics of 
many astrophysical and cosmological flows.

% inverse cascade in astro and cosmology
Without an energy input, turbulence and magnetic fields decay freely
which can be of interest in various astrophysical applications. 
One such example of non-driven MHD turbulence 
is the evolution of magnetic fields in the very early Universe
before the epoch of recombination. 
Scenarios of primordial magnetogenesis include
specific models for inflation \citep[e.g.][]{TurnerWidrow1988} and the cosmological 
phase transitions \citep[e.g.][]{SiglEtAl1997}, but the subsequent evolution of primordial magnetic fields 
is governed by the laws of 
decaying MHD turbulence \footnote{It is worth noting that in a 
flat expanding Universe, the equations 
of relativistic MHD simplify to the nonrelativistic MHD equations if
all quantities are replaced by co-moving quantities and instead of physical time
conformal time is used \cite{Brandenburg:1996fc}.}, \citep[see e.g.][ for a review]{Durrer:2013pga}.
Whether primordial magnetic fields can survive until they might 
e.g.\ serve as seed fields for galactic dynamos or remain as significant 
relics in present-day cosmic voids, depends on the prospect of 
transferring magnetic energy from
small to large spatial scales. 
Such a transfer of magnetic energy is known as inverse cascade and 
is well studied
within MHD turbulence \citep{PouquetFrischLeorat1976,CHB01}.

% inverse cascade in general
The scaling laws of the inverse cascade depend crucially on  
the magnetic helicity $\int_{V}\mathbf{A}\cdot\mathbf{B}~\mathrm{d}V$ which is 
a topological property of the magnetic field.
Here, the volume integral is taken over the product of the vector potential 
$\mathbf{A}$ and the magnetic field $\mathbf{B} = \nabla \times \mathbf{A}$.
Since $\int_{V}\mathbf{A}\cdot\mathbf{B}~\mathrm{d}V$ is a conserved 
quantity in ideal MHD, a decay of the magnetic field necessarily results
into an increase of its correlation length $\ell$. 
Using a phenomenological approach, similar to the Kolmogorov
theory of nonmagnetized turbulence, magnetic helicity conservation
implies scalings with time $t$ of $\ell \propto t^{2/3}$ and $B \propto t^{-1/3}$ 
\citep{BM99PhRvL,MuellerBiskamp2000}.
It has been demonstrated in three-dimensional numerical simulations
that inverse transfer of energy occurs even in
nonhelical decaying MHD turbulence, however with 
less efficiency \citep{BKT15,ReppinBanerjee2017}.
The scaling laws of decaying MHD turbulence, estimated from
phenomenological arguments or extracted from numerical simulations, have been 
applied, for instance, to the evolution of primordial magnetic fields from their
generation until recombination
\citep{KTBN13,BKMPTV17}.

% Extension of MHD to chiral MHD
The assumption of conserved magnetic helicity, however, 
breaks down when charged fermions can be considered as being massless. 
This is the case at high energies that are reached 
in the very early Universe.
In fact, above approximately $10$ MeV \citep{BFR12}, MHD necessarily needs to be 
extended to chiral MHD where only
the sum of magnetic helicity and fermionic chirality
is conserved. 
Fermionic chirality is the divergence of the chiral current 
$j_5$ which is not conserved due to the chiral anomaly, a pure 
quantum effect with macroscopic consequences. 
It can be quantified by the chiral chemical
potential $\mu_5 \equiv \mu_\mathrm{L} - \mu_\mathrm{R}$
with $\mu_\mathrm{R}$ and $\mu_\mathrm{L}$ are the chemical
potentials of right- and left-handed fermions, respectively.
A non-zero $\mu_5$ in the presence of a magnetic field, 
leads to the chiral magnetic effect (CME) \cite{Vilenkin:80a}
which is a macroscopic quantum effect within the
standard model of particle physics and implies an additional electric current along a magnetic
field \citep{SonSurowka2009}.
The CME leads to a magnetic field instability \citep{Joyce:97} which is the subject 
of many studies
\citep{Frohlich:2000en,Frohlich:2002fg,Semikoz:04a,Semikoz:2009ye,
Boyarsky:12a,Semikoz:12a,Tashiro:2012mf,Dvornikov:2011ey,Dvornikov:2012rk,Dvornikov:2013bca,
Dvornikov:2016jth,Manuel:2015zpa,Gorbar:2016klv,Pavlovic:2016mxq,Pavlovic:2016gac,RogachevskiiEtAl2017,FigueroaEtAl2019}.
Recently, the nonlinear dynamics of a chiral plasma was studied also in 
direct numerical simulations (DNS) with a focus on chiral 
dynamos \citep{BSRKBFRK17, SRBBFRK17, MasadaEtAl2018}.
The energies necessary for chiral effects are reached in the 
early Universe and in protoneutron stars, but also in 
heavy ion collisions \citep{Kharzeev:07}. 
Furthermore, chiral MHD is relevant for modeling 
the dynamics of electronic solid-state materials like Weyl semimetals
\citep{GalitskiEtAl2018}.

% How does a magnetic field decay in chiral MHD?
The extension of MHD to chiral MHD raises the following 
questions: How is the inverse cascade affected by 
the new degree of freedom, the chiral chemcial potential?
How much chiral asymmetry can be generated from an
initial helical magnetic field? 
This scenario has been explored by \citet{HironoEtAl2015} who
considered a plasma composed of charged fermions 
with initial magnetic helicity and vanishing chiral asymmetry. 
Within their assumption of negligible velocity fields, they have
identified a three-stage evolution: First, the magnetic helicity
is transferred to fermionic chirality due to the conservation 
law of chiral MHD. 
Second, the total helicity is dominated by fermionic chirality
which eventually leads to a
CME-assisted inverse cascade of magnetic helicity. 
And third, at late times, \citet{HironoEtAl2015} report a self-similar 
evolution of $\mu_5$ and the peak of the magnetic energy spectrum
proportional to $t^{-1/2}$.
This self-similar evolution during the decay of a large $\mu_5$
has been observed in lattice simulations \cite{FigueroaEtAl2019,MaceEtAl2019}.
A remaining open question is, however, how such an evolution 
of chiral MHD is modified in presence of turbulence, where
there can be a strong coupling between the magnetic field 
and the velocity field. 
Understanding decaying chiral helical MHD turbulence
and its differences to the classical MHD scenario is the 
goal of the present study. 

To this end, we investigate the evolution of the magnetic field $\bm{B}$, 
the velocity field $\bm{U}$ and the chiral chemical potential $\mu_5$ 
in both the laminar and the turbulent regime. Our initial conditions are a vanishing 
 chiral asymmetry and velocity field, $\bm{U}=\mu_5=0$ 
and a maximally helical magnetic field. 
These initial conditions are realized, for instance, 
in various inflationary magnetogenesis models
where a pseudo-scalar field generates magnetic fields in a parity violating manner \citep{FieldCarroll2000, AnberSorbo2006,Durrer:2010mq,Caprini:2014mja,Fujita:2015iga,Adshead:2016iae,Fujita:2019pmi}, but does not introduce a chemical potential for fermions.
Note that if the fermion masses are negligible, not only the helical
magnetic field but also the chiral asymmetry can be generated during inflation 
such that the net helicity plus chirality is conserved 
precisely due to the chiral anomaly~\cite{Domcke:2018eki}. 
This alternative initial condition is beyond the scope of the present study but a
target of our future work.

% what we do here:
The paper is structured as follows.
In Section~\ref{sec_background} we briefly review the inverse cascade
in classical MHD and the system of equations in chiral MHD and
introduce our numerical methods.  
In Section~\ref{sec_lowRe} we present an analytical 
derivation of the self-similar inverse cascade in chiral MHD
with a vanishing velocity field and confirm the validity of
our analytical results with three-dimensional numerical simulations.
The transition from a system with vanishing velocity field
to a regime where turbulence is driven efficiently via the
Lorentz force exerted by the helical magnetic field, 
is presented in Section~\ref{sec_highRe}. 
For the limit of large Reynolds numbers, we use a phenomenological
approach to find solutions for the evolution of $\mu_5$. 
The analytical solutions are compared to results from turbulent
DNS. 
We draw our conclusions in Section~\ref{sec_conclusion}.
%

%%%%%%%%%%%%%
% Section 2 %
%%%%%%%%%%%%%

\section{Theoretical background and methods}
\label{sec_background}

\subsection{Review of the classical inverse cascade} 
\label{sec_MHDeqns}

The dynamics of magnetized fluids in the one-fluid 
magnetohydrodynamical limit is described by the following set of equations: 
\begin{eqnarray}
  \frac{\partial \BB}{\partial t} &=& 
     \nabla  \times   \left[{\UU}  \times   {\BB}
      - \eta \, \left(\nab   \times   {\BB}  \right) \right], 
\label{eq_classicind}\\
  \rho \frac{D \UU}{D t} &=& (\nab   \times   {\BB})  \times   \BB
     -\nab  p + \nab  {\bm \cdot} (2\nu \rho \SSSS),
\label{eq_classicNS}\\
  \frac{D \rho}{D t} &=& - \rho \, \nab  \cdot \UU.
\label{eq_classicrho}
\end{eqnarray}
Here, $\BB$ is the magnetic field, $t$ is time, $\UU$ is
the velocity field, and $\rho$ is the mass density.
Furthermore, $p$ is the hydrodynamic pressure, 
${\sf S}_{ij}=1/2(U_{i,j}+U_{j,i})-1/3~\delta_{ij} {\bm \nabla}
{\bm \cdot} \UU$
are the components of the tracefree strain tensor $\SSSS$, 
where commas denote partial spatial derivatives,
and $D/D t = \partial/\partial t + \UU \cdot \nab$ is the
advective derivative.
The Ohmic resistivity is denoted by $\eta$ and 
$\nu$ is the viscosity.
The set of equations is closed by an isothermal equation of 
state, meaning that the pressure is related to
the density via $p=c_{\rm s}^2\rho$, where $c_{\rm s}$ is the 
sound speed.

An important role for the evolution of magnetic
fields is played by magnetic helicity which is defined as
$\int_V \mathcal{H} ~ \mathrm{d}V $ 
with $\mathcal{H}\equiv \AAA \cdot \BB$.
The integral is taken over a periodic volume $V$ or 
over an unbounded volume with the fields falling
off sufficiently rapidly at spatial infinity so that a boundary term can be neglected. 
In these cases, magnetic helicity is gauge invariant.
Its evolution equation can be 
derived by multiplying Faraday's law with its uncurled 
version for the vector potential and yields
\begin{eqnarray}
  \frac{\mathrm{d}}{\mathrm{d}t} \int_V \mathcal{H} ~ \mathrm{d}V
  %\int_V \AAA \cdot \BB ~ \mathrm{d}V 
  = - 2 \eta \int_V \JJ \cdot \BB ~ \mathrm{d}V. 
\label{eq_AB}
\end{eqnarray}
A remarkable consequence of magnetic helicity conservation 
at $\eta \rightarrow 0$ (faster than the current helicity
$\int_V \JJ \cdot \BB ~ \mathrm{d}V$ may possibly diverge)
is the inverse cascade of energy for a 
fully helical magnetic field 
\citep{FrischEtAl1975, PouquetFrischLeorat1976}.

The highly nonlinear evolution of helical decaying MHD turbulence
has been studied intensely with DNS. 
For incompressible 3D magnetohydrodynamic turbulence 
at relatively high $\Rm$, the energy decay 
\citep{BM99PhRvL} as well as scaling relations of the 
energy power spectrum have been analyzed \citep{MuellerBiskamp2000}.
The role of magnetic helicity in the inverse cascade was investigated 
by \citet{CHB01}.
With their DNS, \citet{CHB01} found evidence for a self-similar evolution of magnetic 
energy spectrum with a development of a power law of roughly $k^{-2.5}$ beyond the peak and 
analyzed decay laws for both the kinematic and magnetic energy.
The scaling relations of a helical magnetic field can, in the 
limit of high Reynolds numbers, be derived by using a Kolmogorov-type 
phenomenological approach \citep[see e.g.][]{Biskamp2003}. 
In particular, the magnetic energy evolves as
$\langle \BB^2 \rangle/2 \propto t^{-2/3}$ and the correlation length of
the magnetic field as $\xi \propto t^{2/3}$ 
which has been confirmed by DNS \citep[e.g.][]{KTBN13}.
An inverse transfer of magnetic energy has also been
found for nonhelical magnetic fields, however, it is less
efficient than in the fully helical case 
\citep{KTBN13, Zrake2014,BereraLinkmann2014,BKT15,ReppinBanerjee2017}.

\begin{table*}
\centering
\caption{Summary of all runs presented in this paper. 
The reference runs for laminar (R1 and R1mhd) and turbulent (R8 and R8mhd) simulations which 
are presented in detail in Figures~\ref{fig_reference_runs_laminar} and \ref{fig_reference_runs_turbulent}, 
respectively, are highlighted by
bold font. 
The amplitude of the initial magnetic power spectrum is exactly the same for 
all runs R1--R7. 
Runs R7 and R7mhd have, however, a larger initial rms magnetic field strength $B_0$
which is due to the higher resolution, meaning the larger number of modes available 
in R7 (the exponentially suppressed tail extends to higher wave numbers). 
For runs R8, R8b, and R8mhd, a larger initial amplitude has been set. 
}
\begin{tabular}{| l  l  l | l  l  l l l | l l l | } 
 \hline
                   &        &              & \hspace{-.01cm}Input parameters:\hspace{-1.2cm} &  &  &  &  &   \hspace{-.01cm}Measured parameters:\hspace{-1.2cm}  &    &   \\  
   Name            & MHD    & resolution   & $\dfrac{B_{\mathrm{rms},0}}{10^{-2}}$ & $k_{\mathrm{p},0}$ & $\mu_{5,0}$ & $\dfrac{B_{\mathrm{rms},0}}{\eta}$ & $\dfrac{\lambda B_{\mathrm{rms},0}^2}{k_{\mathrm{p},0}}$  &   $\RmABmin$  &   $\Rmkpkone$ &  $\Rmmax$  \\  
%   Name            & MHD    & resolution   & $10^2\dfrac{ B_{\mathrm{rms},0}}{c_\mathrm{s}}$ & $\dfrac{k_{\mathrm{p},0}}{k_1}$ & $\dfrac{\mu_{5,0}}{k_1}$ & $\dfrac{B_{\mathrm{rms},0}}{\eta k_1}$ & $\dfrac{\lambda B_{\mathrm{rms},0}^2}{k_1 k_{\mathrm{p},0}}$  &   $\RmABmin$  &   $\Rmkpkone$ &  $\Rmmax$  \\  
 \hline 
   R1a            & chiral             & $320^3$      & $1.153$      & $85$      & 0         & $11.53$         & $1.662$       & $7.2\times10^{-6}$     & $1.9\times10^{-5}$      & $9.6\times10^{-4}$  \\
   \textbf{R1}    & \textbf{chiral}    & $\bf 320^3$  & $\bf 1.153$  & $\bf 85$  & $\bf 0$   & $\bf 11.53$     & $\bf 16.618$    & $\bf 2.7\times10^{-5}$ & $\bf 1.4\times10^{-1}$  & $\bf 1.6\times10^{-1}$  \\
   R1b            & chiral             & $320^3$      & $1.153$      & $85$      & 0         & $11.53$         & $166.176$        & $1.6\times10^{-3}$     & $2.3\times10^{-1}$       & $2.3\times10^{-1}$  \\
   \textbf{R1mhd} & \textbf{classic} & $\bf 320^3$  & $\bf 1.153$  & $\bf 85$  & $\bf -$   & $\bf 11.53$     & $\bf -$        & $ \bf 2.4\times10^{-6}$     & $ \bf2.9\times10^{-5}$      & $ \bf1.0\times10^{-3}$   \\
   R2             & chiral             & $320^3$      & $1.153$      & $85$      & 0         & $23.06$         & $16.618$        & $1.2\times10^{-4}$     & $-$                     & $2.3\times10^{-1}$   \\
   R3             & chiral             & $320^3$      & $1.153$      & $85$      & 0         & $115.3$        & $16.618$        & $2.6\times10^{-3}$     & $3.4\times10^{0}$       & $3.6\times10^{0}$   \\
   R4             & chiral             & $320^3$      & $1.153$      & $85$      & 0         & $230.6$        & $16.618$        & $9.8\times10^{-3}$     & $7.1\times10^{0}$       & $7.1\times10^{0}$   \\
   R5             & chiral             & $320^3$      & $1.153$      & $85$      & 0         & $576.5$        & $16.618$        & $8.1\times10^{-2}$     & $-$                     & $6.6\times10^{0}$   \\
   R6             & chiral             & $320^3$      & $1.153$      & $85$      & 0         & $1153.0$       & $16.618$        & $1.6\times10^{0}$      & $-$                   & $9.9\times10^{0}$   \\
   R7             & chiral             & $512^3$      & $1.400$      & $85$      & 0         & $2800.0$       & $24.5$        & $1.4\times10^{1}$      & $-$                     & $2.5\times10^{1}$    \\
   R7mhd          & classic          & $512^3$      & $1.400$      & $85$      & $-$       & $2800.0$       & $-$            & $8.3\times10^{0}$      & $-$                     & $1.2\times10^{1}$   \\
   \textbf{R8}    & \textbf{chiral}    & $\bf 512^3$  & $\bf 4.667$  & $\bf 85$  & $\bf 0$   & $\bf 9333.6$   & $\bf 24.501$    & $\bf 1.1\times10^{2}$  & $\bf 3.2\times10^{2}$   & $\bf 3.2\times10^{2}$  \\
   R8b            & chiral            & $512^3$  & $4.667$  & $ 85$  & $ 0$   & $9333.6$   & $2450.140$    & $5.5\times10^{1}$  & $-$   & $7.8\times10^{2}$  \\
   \textbf{R8mhd} & \textbf{classic} & $\bf 512^3$  & $\bf 4.667$  & $\bf 85$  & $\bf -$   & $\bf 9333.6$   & $\bf -$        & $\bf 2.6\times10^{2}$  & $\bf 2.7\times10^{2}$   & $\bf 2.7\times10^{2}$  \\
 \hline
\end{tabular}
\label{tab_DNSoverview}
\end{table*}
%%%%%%%%

\subsection{Chiral MHD equations} 
\label{sec_chiralMHDeqns}

At high energies, MHD necessarily needs to be generalized to chiral MHD in which the chiral
asymmetry appears as a new degree of freedom. 
Here, an asymmetry between the number densities of left- and right-handed 
fermions gives rise to the CME that results
in an electric current proportional to the magnetic field 
and a chiral chemical potential
\begin{eqnarray}
  \mu_5^\mathrm{phys} \equiv \mu_{_{\rm L}} -\mu_{_{\rm R}},
\label{eq_mu5}
\end{eqnarray}
where $\mu_{_{\rm L}}$ and $\mu_{_{\rm R}}$ are the 
chemical potentials of left- and right-handed fermions,
respectively. 
In the following, we will replace $\mu_5^\mathrm{phys}$ by a normalized
version, $\mu_5 \equiv (4 \alphaem /\hbar c) \mu_5^\mathrm{phys}$,
that has the same units as a wave number.

Chiral asymmetry is coupled to magnetic helicity and 
significantly modifies the phenomenology of the plasma.
The set of chiral MHD equations  
is given by \citep{RogachevskiiEtAl2017}:
\begin{eqnarray}
  \frac{\partial \BB}{\partial t} &=& \nab   \times   \left[{\UU}  \times   {\BB}
     - \eta \, \left(\nab   \times   {\BB}
     - \mu_5 {\BB}   \right) \right], 
\label{eq_ind}\\
  \rho \frac{D \UU}{D t} &=& (\nab   \times   {\BB})  \times   \BB
     -\nab  p + \nab  {\bm \cdot} (2\nu \rho \SSSS),
\label{eq_NS}\\
  \frac{D \rho}{D t} &=& - \rho \, \nab  \cdot \UU, 
\label{eq_rho}\\
  \frac{D \mu_5}{D t} &=& \mathscr{D}_5 \, \Delta \mu_5
     + \lambda \, \eta \, \left[{\BB} {\bm \cdot} (\nab   \times   {\BB})
     - \mu_5 {\BB}^2 \right].   
\label{eq_mu5D}
\end{eqnarray}
Here, the 
chiral vortical effect,
the chiral separation effect, and chirality flipping are neglected.
The latter is well justified at sufficiently high 
temperatures while the former might not be a very good 
approximation in the case of high Reynolds number where 
large vortical velocities can be generated.
The new equation (\ref{eq_mu5D}) includes a diffusion term with the 
diffusion constant $\mathscr{D}_5$ that is only relevant when
$\mu_5$ is strongly varying in space. 
In this study, the evolution of $\mu_5$ will be mostly affected by 
the electromagnetic field via
the second term on the right-hand side of Equation (\ref{eq_mu5D}).
The strength of the feedback
is controlled by the chiral feedback parameter 
\begin{eqnarray}
  \lambda = 3 \hbar c \left(\frac{8 \alphaem}{ \kB T} \right)^2,
\label{eq_lambda}
\end{eqnarray}
which is valid for $\kB T \gg \max(|\mu_L|,|\mu_R|$) \citep{BFR15}.
In the following, $\lambda$ will be considered constant,
yet one should keep in mind that it scales with temperature 
$T^{-2}$ \citep[see, e.g.][]{BFR12}.

The system of Equations~(\ref{eq_ind})--(\ref{eq_mu5D}) implies a conservation law:
\begin{align}
 &\frac{\partial }{\partial t} \left(\frac{\lambda }{ 2}  \mathcal{H}
 %{\bm A} {\bm \cdot} \BB
  + \mu_5 \right) 
  \nonumber\\
  &+ \nab  {\bm \cdot} \left(\frac{\lambda}{ 2} \left({\bm \EE} \times   {\bm A} + \BB \, \Phi\right)  -  \mathscr{D}_5 \nab  \mu_5\right)  
= 0,
\label{eq_CL}
\end{align}
where 
${\bm \EE} = - c^{-1} \, [ {\bm \UU} {\bm \times} {\BB} + \eta \, (\mu_5 {\BB} - {\bm \nabla} {\bm \times} {\BB} ) ]$ 
is the electric field and $\Phi$ is the
electrostatic potential.
Hence, the total chirality $\langle \mathcal{H} \rangle + 2 \langle \mu_5 \rangle/\lambda$
is a conserved quantity,
where $\langle \mu_5 \rangle$
is the mean value of the chiral chemical potential and
$\langle \mathcal{H}\rangle \equiv V^{-1}\int \AAA \cdot \BB \, dV$
is the mean magnetic helicity density in the volume $V$.

\subsection{Numerical methods} 
\label{sec_DNSmethods}

%general
To go beyond the limitations of analytical calculations, we
use the \textsc{Pencil Code}~\footnote{\textit{http://pencil-code.nordita.org/}} for
solving Equations~(\ref{eq_classicind})--(\ref{eq_classicrho}) for 
classical MHD and Equations~(\ref{eq_ind})--(\ref{eq_mu5D}) for 
chiral MHD, respectively.
The system of equations is solved 
in a three-dimensional periodic domain of size $L^3 = (2\pi)^3$ 
via a third-order
accurate time-stepping method of \cite{Wil80} and sixth-order explicit finite differences
in space \citep{BD02,Bra03}.
The time step is specified as the Courant time step, however, for 
our reference runs we initially use very small manually set 
time steps to resolve the very early time evolution in more detail. 
After the initial phase, the automatic Courant time step
is used in the reference runs. 
The resolution is varied between $320^3$ and $512^3$.
%normalization and code units
The smallest wave number covered in the numerical domain is 
$k_1 = 2\pi/L = 1$ which we use as
normalization of length scales. 
All velocities are normalized to the sound speed $\cs = 1$ and the
mean fluid density to $\meanrho = 1$.
Time is normalized by the diffusion time $t_\eta = \eta^{-1}$,
where $\eta$ is the Ohmic resistivity.

%initial conditions
In this study, all runs are initialized with vanishing chiral chemical 
potential $\mu_5$ and a strong helical random magnetic 
field~\footnote{For the setting the initial condition
of the helical random magnetic fields, we use the routine 
\textit{power\_randomphase\_hel} in the \textsc{Pencil Code}.}.
In practice, the magnetic field is 
set up via the vector potential $\AAA(\boldsymbol{x})$ which is constructed 
from a random and $\delta$-correlated three-dimensional vector field in 
real space.
The magnetic field is calculated from the Fourier transform of
$\AAA(\boldsymbol{x})$ via $\BB(\boldsymbol{k}) = i \boldsymbol{k} \times \AAA(\boldsymbol{k})$.
Then the magnetic field is scaled by functions of $k$ to adjust
the shape of the magnetic energy spectrum $E_{\rm M}(k)=2\pi B^2(k)k^2$ for which
we use a slope proportional to $k^4$ between $1$ and 
the initial wave number of the energy-carrying eddies 
$k_{\mathrm{p},0}=85$, i.e.\ the initial peak of the magnetic energy spectrum.
For $k>k_{\mathrm{p},0}$ in all runs, the spectrum is 
suppressed.
Multiplication by the operator $P_{ij}(\boldsymbol{k}) - i \sigma_\mathrm{M} \hat{k}_l$,
where $P_{ij} = \delta_{ij} - \hat{k_i} \hat{k_j}$ is the projection operator,
ensures a fully helical magnetic field for $\sigma_\mathrm{M}=\pm 1$.
Finally, the energy spectrum is normalized such that
$\bra{\BB^2}/2=\BB_\mathrm{rms}^2/2=\int_{1}^{k_\mathrm{max}} E_{\rm M}(k)\,\mathrm{d}k$,
where the integration is performed over the entire numerical domain, i.e.\ from $k=1$ up to
the maximally resolved wave number $k_\mathrm{max}$.
We note that these initial conditions of the magnetic field 
are similar to the ones used in previous studies of 
decaying MHD turbulence, like in \citet{BKMPTV17}.

% Turbulence
No external forcing is applied to drive turbulence in our simulations, i.e.,
the velocity field is purely driven via the Lorentz force that is exerted on
the flow through the magnetic field. 
The transition to a turbulent plasma occurs when the
magnetic and kinetic Reynolds numbers,
$\Rm = u_\mathrm{rms}/(k_\mathrm{f} \eta)$ and 
$\Rey = u_\mathrm{rms}/(k_\mathrm{f} \nu)$, 
respectively, become much larger than unity.
Here, $u_\mathrm{rms}$ is the rms velocity and $k_\mathrm{f}$ is the 
wave number on which kinetic energy is injected in the system.
For magnetically driven turbulence, $k_\mathrm{f}$ 
corresponds to the inverse correlation length of the magnetic field
and we will use $k_\mathrm{f} = k_\mathrm{p}(t)$.  
Viscosity $\nu$ and Ohmic resistivity $\eta$ are implemented 
explicitly in the code. 
To explore systems with different Reynolds numbers, the
values of $\nu$ and $\eta$ are systematically changed,
while
their ratio, i.e.\ the magnetic Prandtl number $\Pm$, 
is set to unity for all of the simulations. 
We note, that the choice of $\Pm=1$ does not 
reflect the situation in most astrophysical applications. 
However, $\Pm\ll1$ or $\Pm\gg1$ are notoriously difficult 
to treat in DNS, since that requires a large separation of scales.
In what follows, we therefore only mention the magnetic Reynolds 
number which in our settings equals the kinetic one.

%overview of all DNS
An overview of the input parameters and characteristic numbers of
all runs discussed in this work is presented in
Table~\ref{tab_DNSoverview}.

%%%%%%%%%%%%%
% Section 3 %
%%%%%%%%%%%%%

\section{Inverse cascade in chiral MHD with a vanishing velocity field}
\label{sec_lowRe}

In this section, we discuss the evolution of a decaying helical magnetic field
for simulations where the velocity field can be neglected throughout the entire
simulation time. 
We note, however, that Equation~(\ref{eq_NS}) is nevertheless included
in the DNS.

\begin{figure*}
\centering
% /Users/jennifer/Science/Coding/pencil-code/jenny/chiral_fluids/turbulent_decay/320_3D_kf80_B3em3_eta1e-3_hel_baobab/postproc/time_evolution.pdf -> ./figures/time_evolution__linearMHD.pdf
% /Users/jennifer/Science/Coding/pencil-code/jenny/chiral_fluids/turbulent_decay/320_3D_kf80_mu00_lambda1e7_B3em3_eta1e-3_hel_baobab/postproc/time_evolution.pdf  -> ./figures/time_evolution__linearchiralMHD.pdf
% /Users/jennifer/Science/Coding/pencil-code/jenny/chiral_fluids/turbulent_decay/320_3D_kf80_B3em3_eta1e-3_hel_baobab/postproc/spec_mag.pdf -> ./figures/spec_mag__linearMHD.pdf
% /Users/jennifer/Science/Coding/pencil-code/jenny/chiral_fluids/turbulent_decay/320_3D_kf80_mu00_lambda1e7_B3em3_eta1e-3_hel_baobab/postproc/spec_mag.pdf -> ./figures/spec_mag__linearchiralMHD.pdf
  \includegraphics[width=0.96\textwidth]{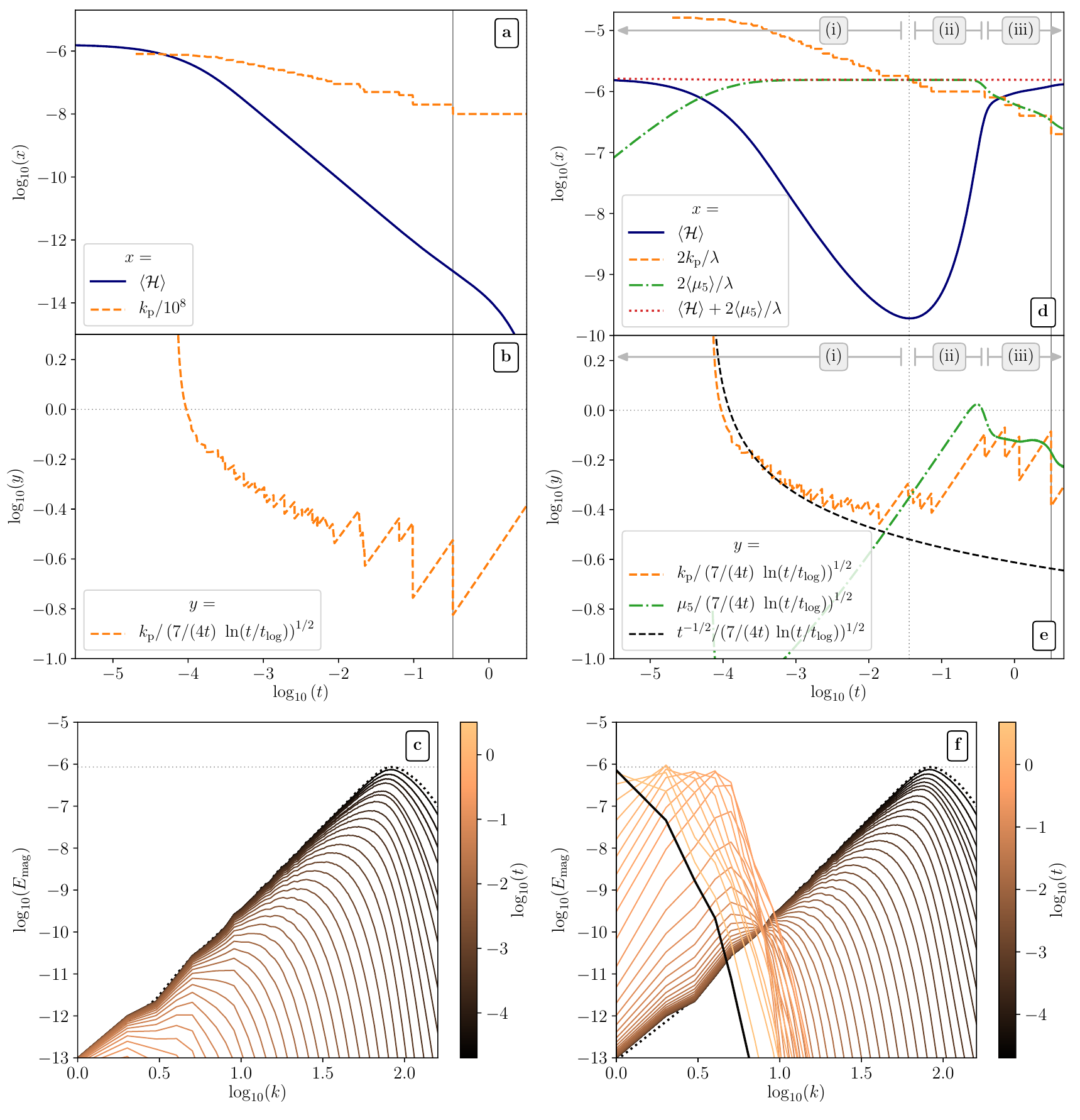}
\caption{Comparing a run of a decaying helical magnetic field
in classical MHD (\textit{Left}, run R1mhd) 
with chiral MHD (\textit{Right}, run R1). 
The velocity field in these simulations is negligible.
% The runs have high diffusivity and can be considered as being 
% laminar during the entire simulation time.
\textit{Top:} Time evolution of the mean magnetic helicity 
$\langle\mathcal{H} \rangle$ and the wave number on 
which the magnetic energy spectrum
has its maximum $k_\mathrm{p}$. 
The time $t$ is 
normalized to the resistive time $t_\eta$.
For chiral MHD, we also show the mean 
chiral chemical potential $\langle\mu_5\rangle$ over the
chiral feedback parameter $\lambda$ as well as the conserved quantity 
$\langle\mathcal{H} \rangle + 2\langle\mu_5\rangle/\lambda$.
The solid vertical lines indicate the
time when $k_\mathrm{p}$ has reached the 
minimum wave number possible in the numerical domain, $k = 1$,
and the dotted vertical lines indicates the time
at which the minimum of magnetic
helicity is reached.
\textit{Middle:} $k_\mathrm{p}$ and $\langle\mu_5\rangle$ normalized
by their theoretically expected scaling in the self-similar 
evolution phase (see Section~\ref{sec_analytical}). 
\textit{Bottom:} Evolution of the magnetic energy spectrum. 
The thick dotted black 
lines
show the initial spectra and the 
thick solid black lines show the final spectra of the simulations.
}
\label{fig_reference_runs_laminar}
\end{figure*}

\subsection{Analytical solutions in the limit of dynamically insignificant velocity fields}
\label{sec_analytical}

Provided that helical magnetic fields are nonzero but the chiral chemical potential is 
vanishing $\mu_5(t_0)\equiv \mu_{5,0}=0$ at the initial time, the second term in the right-hand side in Equation~\eqref{eq_mu5D} sources $\mu_5$.
$|\mu_5|$ may grow until it reaches the maximum value at which the second term and the third term  cancel each other, 
\begin{equation}
|\mu_5(t)| \simeq \left|\frac{\bm B\cdot(\nabla \times \bm B)}{B^2}\right|\simeq k_p(t),
\end{equation}
where we ignore the diffusion term $\mathscr{D}_5 \Delta \mu_5$ in Equation~\eqref{eq_mu5D}. 
Due to the conservation law $\langle\mathcal{H}\rangle +2\langle\mu_5\rangle/\lambda= \mathrm{const.}$, however, 
$\mu_5$ cannot exceed the initial value of the conserved quantity,
\begin{equation}
|\mu_5(t)|\lesssim \lambda |\mathcal{H}(t_0)| \simeq \lambda B^2_0/k_{\mathrm{p},0},
\end{equation}
where $B_0$ is the strength of the initial magnetic field with maximum energy on
the wave number $k_{\mathrm{p},0}$.
Therefore, the chiral chemical potential generated from
magnetic helicity is given by
\begin{eqnarray}
  |\mu_{5}(t)| \simeq \mathrm{min}\left(k_\mathrm{p}(t), \lambda B_0^2/k_{\mathrm{p},0}\right).
\label{eq_mu5max}
\end{eqnarray}
Since $k_{\mathrm{p}}(t)$ decreases in time 
due to the diffusion of magnetic energy on large $k$ (small scales),  even if 
$\mu_5$ reaches $\lambda B_0^2/k_{\mathrm{p},0}$ at early times, it starts decaying when
$k_\mathrm{p}(t)$ becomes smaller than $\mu_5$
because of the third term in Equation~\eqref{eq_mu5D}.
Thus $\mu_5(t)$ eventually follows $k_{\mathrm{p}}(t)$.
Depending on the initial conditions, two different scenarios can be
identified for decaying chiral MHD, as long as the velocity
field can be neglected.

If $k_{\mathrm{p},0} > \lambda B_0^2/k_{\mathrm{p},0}$ the system
evolves in three phases: \\
(i) Production of $\langle\mu_5\rangle$ at the expense of magnetic helicity $\langle\mathcal{H}\rangle$, with an efficiency depending on $\lambda$ up to $\lambda B_0^2/k_{\mathrm{p},0}$. \\
%The production terminates when the conserved quantity is dominated by $\mu_5$.
(ii) Once $k_\mathrm{p}(t)$ has decayed to $\langle\mu_5\rangle \simeq \lambda B_0^2/k_{\mathrm{p},0}$, a chiral dynamo,
the so-called $v_\mu^2$ dynamo as discussed in \cite{RogachevskiiEtAl2017},
leads to an exponentially fast restoration 
of $\langle\mathcal{H}\rangle$.\\
(iii) A self-similar inverse cascade sets in with $\kp\approx \langle\mu_5\rangle$, during which $\langle\mu_5\rangle$
is converted into $\langle\mathcal{H}\rangle$
according to the conservation law.\\
In the other case with $k_{\mathrm{p},0} < \lambda B_0^2/k_{\mathrm{p},0}$, 
the growth of $\langle\mu_5\rangle$ stops when it becomes comparable to $k_{\mathrm{p}}(t)$ 
during the phase (i) and the system immediately enters the phase (iii) by skipping (ii). 
In this case, the magnetic helicity always dominates the conserved quantity and the evolution
of magnetic fields is not significantly altered by $\langle\mu_5\rangle$.

As we show in Appendix~\ref{sec_SSEderivation} that during this self-similar evolution in the phase (iii), $\mu_5$ and $k_\mathrm{p}$ evolve as
\begin{eqnarray}
  |\mu_{5}(t)| \approx 
  k_\mathrm{p}(t) \approx \left[\frac{3 + n}{4 \eta t}~\ln\left(\frac{t}{t_\mathrm{log}}\right)\right]^{1/2},
\label{eq_SSE}
\end{eqnarray}
where $n$ denotes the slope of the initial magnetic helicity spectrum
$\langle\mathcal{H}(t_0) \rangle_k\propto k^n$. 
Our DNS are initiated with $n=4$.
We found that the logarithmic correction time is roughly given by 
$t_\mathrm{log}=(2 k_{\mathrm{p},0}^2)^{-1}$ in our simulations,
which is written as $t_\mathrm{log} =(2k_{p,0}^2/k_1^2)^{-1} t_\eta$
when using explicit units.
This result up to the logarithmic correction term has also been 
found in~\cite{HironoEtAl2015}.

\subsection{Comparison of classical MHD with a three-phase chiral MHD scenario in DNS}

The reference run for a three phase scenario of a decaying 
magnetic field in chiral MHD is R1.
In Figure~\ref{fig_reference_runs_laminar}, R1 is
compared to a classical MHD analog (R1mhd).
The parameters and initial conditions in both runs are 
the same but in R1 the plasma evolves according to 
Equations~(\ref{eq_ind})--(\ref{eq_mu5D}) and in R1mhd according to
Equations~(\ref{eq_classicind})--(\ref{eq_classicrho}). 
R1 is presented in the right panels of  
Figure~\ref{fig_reference_runs_laminar} and
R1mhd is presented in the left panels.

The time evolution of the mean magnetic helicity, 
$\langle\mathcal{H} \rangle$, and the wave number on 
which the magnetic energy spectrum
has its maximum, $k_\mathrm{p}$, are presented in the top
row of Figure~\ref{fig_reference_runs_laminar}. 
% Time is normalized by the resistive timescale 
% $t_\eta = (\eta k_1^2)^{-1}$.
In the classical nonideal MHD case the magnetic 
helicity decreases by approximately eight orders of magnitude
during one resistive
time, i.e.\ until $t\approx 1$. 
Resistivity acts on small spatial scales, e.g.~large
wave numbers $k$. 
This leads to a decrease of the magnetic energy on large $k$ 
and therefore a move of the peak scale of the magnetic energy
spectrum, $k_\mathrm{p}$, to smaller $k$.
Note, that $k_\mathrm{p}$ has discrete values only, leading to steps in its time evolution that become more evident at late times when $k_\mathrm{p}$ approaches $1$.
Since the velocity field is negligible during the
entire run, there is no inverse transfer of magnetic energy, as 
can be seen in the evolution of the magnetic energy spectrum,
see Figure~\ref{fig_reference_runs_laminar}c. 
The time evolution of $k_\mathrm{p}$, normalized to
the theoretically predicted value for chiral MHD 
given in Equation~(\ref{eq_SSE}), is presented in the middle
row of Figure~\ref{fig_reference_runs_laminar}.
For R1mhd, Equation~(\ref{eq_SSE}) is not valid and
therefore the orange dashed line in Figure~\ref{fig_reference_runs_laminar}b  
moves away from $1$ with increasing time. 

The time evolution of $\meanAB$ 
in R1 is significantly different from the one in classical MHD, 
as can be see in 
Figure~\ref{fig_reference_runs_laminar}d.
First, $\meanAB$ decreases by
roughly two orders of magnitude. 
At the same time, a mean chiral chemical potential 
$\langle\mu_5\rangle$ is generated, such that the
sum $\langle\mathcal{H}\rangle +2\langle\mu_5\rangle/\lambda$
is conserved during the entire run.
The value of $k_\mathrm{p}$ decreases in time in R1, 
but not as quickly as in R1mhd.
The three phases described in Section~\ref{sec_analytical} 
can be clearly distinguished in 
Figure~\ref{fig_reference_runs_laminar}d:
Phase (i) during which $\meanAB$
decreases and which
ends at $t\approx4\times10^{-2}$ is followed by a
phase of dynamo amplification, phase (ii).
For $t\gtrsim 4 \times10^{-1}$,
$\langle\mu_5\rangle$ and $k_\mathrm{p}$ evolve in a
self-similar way, what was defined as phase (iii). 
During this phase, the evolution of $\langle\mu_5\rangle$ and $k_\mathrm{p}$ 
is reasonably well described by Equation~(\ref{eq_SSE}),
as can be seen in 
Figure~\ref{fig_reference_runs_laminar}e.
For comparison with the scaling of $k_\mathrm{p}\propto t^{-1/2}$, 
we have added the black dashed line in 
Figure~\ref{fig_reference_runs_laminar}e from which the 
simulation data clearly deviates in phase (iii).
The time evolution of the magnetic energy spectrum in run R1 
(right bottom panel) 
is very different from the one in R1mhd 
(Figure~\ref{fig_reference_runs_laminar}c).
The main difference occurs at late times,
where in R1mhd, the magnetic energy first
grows on $k \approx 5$ and then moves to smaller wave numbers
in an CME-assisted inverse cascade.

\subsection{Dependence on the chiral feedback parameter $\lambda$}
\begin{figure}
\centering
%/Users/jennifer/Science/Coding/pencil-code/jenny/chiral_fluids/turbulent_decay/postproc_mu50/energy_t__lambda.py
   \includegraphics[width=0.48\textwidth]{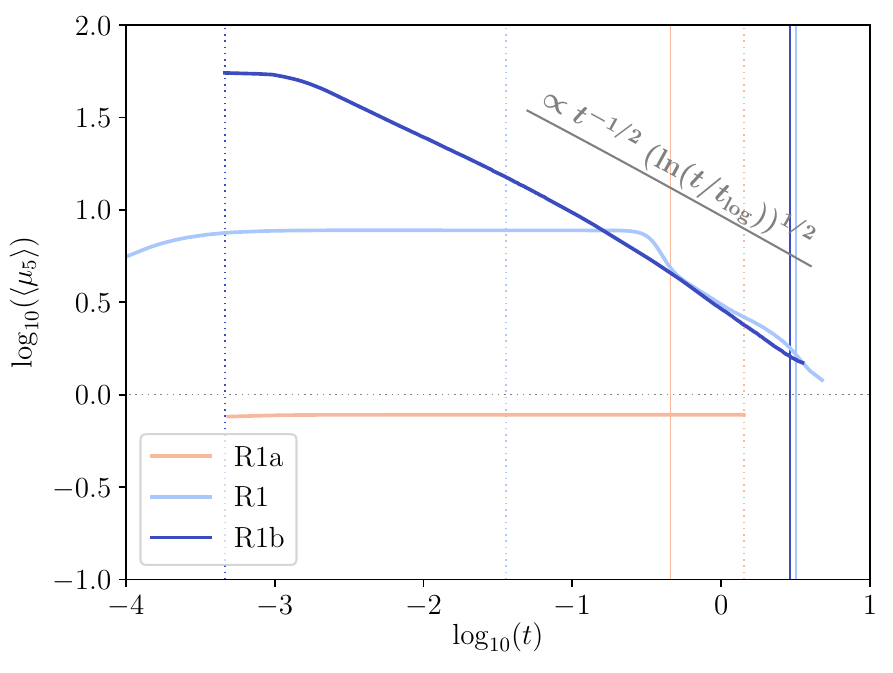} 
\caption{The mean chiral chemical potential, $\langle \mu_5 \rangle$, as
a function of time for runs
R1a, R1, and R1b which differ only in their value of 
$\lambda$ (see Table~\ref{tab_DNSoverview}).
The solid vertical lines indicate the
time when $k_\mathrm{p}$ has reached the 
minimum wave number possible in the numerical domain, $k = 1$,
and the dotted vertical lines indicates the time
at which the minimum of magnetic
helicity is reached.
The horizontal gray dotted line indicates the threshold
for $\mu_5$ above which a dynamo instability occurs
in the numerical box.
}
\label{fig_mu5rms_t__lambda}
\end{figure}

\begin{figure}
\centering
%/Users/jennifer/Science/Coding/pencil-code/jenny/chiral_fluids/turbulent_decay/postproc_mu50/abm_t__lambda.py
   \includegraphics[width=0.48\textwidth]{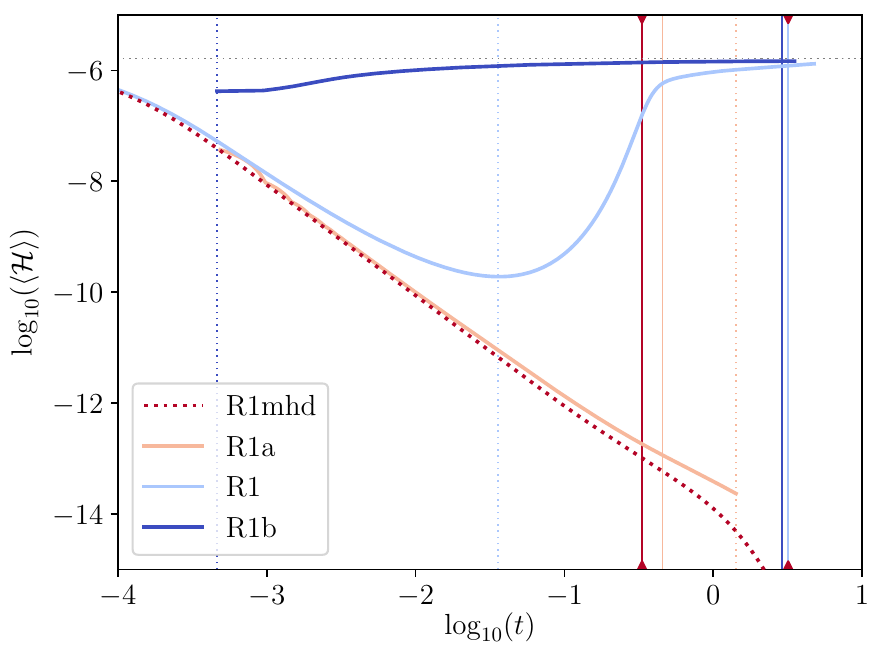} 
\caption{The mean magnetic helicity, $\langle\mathcal{H} \rangle$, as
a function of time for runs
R1mhd, R1a, R1, and R1b (see Table~\ref{tab_DNSoverview}).
The horizontal gray dotted line indicates the initial
value of $\langle\mathcal{H} \rangle$ which is
equal for all runs presented in this figure.
The vertical lines indicate the same characteristic times
as in Figure~\ref{fig_mu5rms_t__lambda}.
}
\label{fig_abm_t__lambda}
\end{figure}

\begin{figure}
\centering
%/Users/jennifer/Science/Coding/pencil-code/jenny/chiral_fluids/turbulent_decay/postproc_mu50/kpcorr_t__lambda.py
   \includegraphics[width=0.48\textwidth]{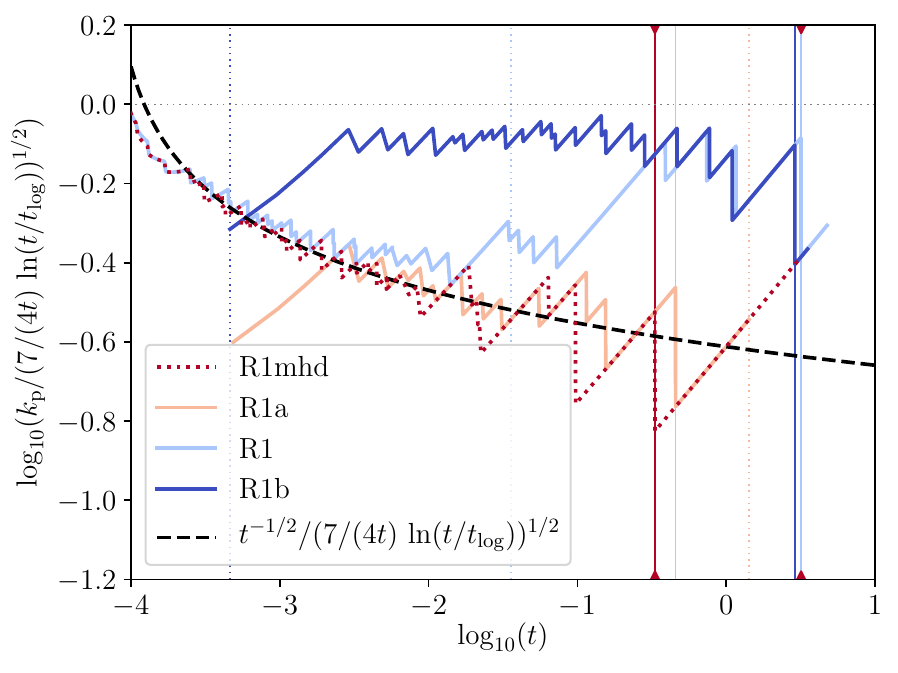} 
\caption{The ratio of the peak scale of the magnetic energy spectrum $k_{\mathrm{p}}$ measured
from the simulation data over the theoretical prediction (see Equation~\ref{eq_SSE})
as a function of time for 
runs R1, R1a, R1b, and R1mhd (see Table~\ref{tab_DNSoverview}). 
The vertical lines indicate the same characteristic times
as in Figure~\ref{fig_mu5rms_t__lambda}.}
\label{fig_scaling_t__lambda}
\end{figure}

The generation of a chiral asymmetry and the subsequent evolution of the
plasma depends strongly on the chiral feedback parameter $\lambda$
or, in dimensionless units, $\lambda B_0^2/k_{\mathrm{p},0}$
as compared to the value of $k_{\mathrm{p},0}$.
As discussed in Section~\ref{sec_analytical}, for 
$\lambda B_0^2/k_{\mathrm{p},0} \ll k_{\mathrm{p},0}$,
we expect a three-phase evolution, while for 
$\lambda B_0^2/k_{\mathrm{p},0} \gg k_{\mathrm{p},0}$,
our models suggest a 2-phase scenario.
In this section, we present a comparison of R1 with a run with 
a smaller value of $\lambda$, R1a, and a run with 
a larger value of $\lambda$, R1b. 

The time evolution of $\mu_5$ in runs R1a, R1, and R1b 
is presented in Figure~\ref{fig_mu5rms_t__lambda}.
The largest maximum value of $\mu_5\approx k_{\mathrm{p},0} = 85$ is 
reached in run R1b, for which 
$k_{\mathrm{p},0} = 85 <\lambda B_0^2/k_{\mathrm{p},0} \approx 166.2$. 
This run, almost instantaneously, enters the self-similar phase
where both $\mu_5$ and $k_\mathrm{p}$ decay proportional
%to $((t/t_\eta)~\mathrm{log}(t/t_\eta))^{-1/2}$
to $t^{-1/2}\left(\mathrm{ln}(t/t_\mathrm{log})\right)^{1/2}$.
This scaling is indicated as a grey line in 
Figure~\ref{fig_mu5rms_t__lambda}.
During the entire run time of R1b, 
magnetic helicity is sourced by $\mu_5$ and, 
therefore, a CME-assisted inverse cascade of magnetic energy occurs. 

The situation is different in the reference run R1,
where $k_{\mathrm{p},0} = 85 > \lambda B_0^2/k_{\mathrm{p},0} \approx 16.6$. 
In R1, a value of $\mu_5 \approx 8$ is generated quickly and stays 
constant up to $t \approx 0.3$. 
At that time the value of $\meanAB$ is up to two orders
of magnitude less than its initial value; see the time evolution 
of $\meanAB$ in Figure~\ref{fig_abm_t__lambda}. 
Via a chiral dynamo $\meanAB$ grows exponentially in time
around $t \approx 0.2$. 
Subsequently, R1 enters the self-similar evolution
phase with an CME-assisted inverse cascade.

Run R1a has the lowest value of $\lambda$ where
$k_{\mathrm{p},0} = 85 > \lambda B_0^2/k_{\mathrm{p},0} \approx 1.7$. 
For these parameters, a maximum value of $\mu_5\approx 0.8$ is
generated. 
For this run, unlike in our reference runs, e.g.~R1, we have applied an automatic time stepping method during the entire simulation time. 
As a result, the value of $\mu_5$ increases from $0$ to $\approx 0.8$ within the first timestep, leaving the $\mu_5$ generation phase unresolved.
With the maximum $\mu_5$ being less than $1$, there can be no chiral dynamo
instability within the simulation domain. 
Therefore, after its initial generation,
$\mu_5$ remains approximately constant 
throughout the entire simulation time. 
Simultaneously, $\meanAB$ decreases in time as can be seen in
Figure~\ref{fig_abm_t__lambda}.
For comparison, also the MHD run R1mhd is presented
Figure~\ref{fig_abm_t__lambda}. 
Here, the value of $\meanAB$ decreases constantly
at a rate that is only slightly larger than for 
R1a. 
At $t \approx 0.45$, the peak of the magnetic 
energy spectrum in R1a reaches the minimum wave number
in the simulation domain, $k_\mathrm{p}=1$.
A dynamo instability for $\mu_5= 0.8$ would occur at
$k_5 = \mu_5/2 = 0.4$. 
Since this is outside of the numerical domain, a 
dynamo and a subsequent CME-assisted inverse cascade
is not seen in R1a. 
We stress, however, that this is purely caused by the
finiteness of the numerical domain. 
For infinite systems, a three-phase scenario is 
expected for all high-energy plasmas with parameters such that
$k_{\mathrm{p},0} > \lambda B_0^2/k_{\mathrm{p},0}$.

The time evolution of the peak scale of the magnetic energy
spectrum $k_\mathrm{p}$ in R1, R1a, R1b, and R1mhd is presented in
Figure~\ref{fig_scaling_t__lambda}.
Here, $k_\mathrm{p}$ is normalized by the analytical solution
in the self-similar phase, Equation~(\ref{eq_SSE}).
For R1b, $k_\mathrm{p}/\left[7/(4 t)  \, \ln(t/t_\mathrm{log})\right]^{1/2}$
has a constant value of $\approx 0.9$ for $t \gtrsim 3\times10^{-3} $
which is equivalent to a few timesteps of the simulation. 
Hence, Equation~(\ref{eq_SSE}) describes the evolution 
during the CME-assisted inverse cascade well.
We show the direct comparison with the scaling $\kp \propto t^{-1/2}$
as the black dashed line in Figure~\ref{fig_scaling_t__lambda}.
The deviation from the $\kp \propto t^{-1/2}$ scaling is clearly visible
in our simulations once they have entered phase (iii) in which $\kp$ and $\langle\mu_5 \rangle$
evolve self-similarly.
This is strong evidence for the need of a logarithmic correction that emerges 
naturally
in our analytic derivation that is given in the appendix.

For the classical MHD simulation, R1mhd, 
$k_\mathrm{p}/(7/(4 t)  \, \ln(t/t_\mathrm{log}))^{1/2}$
is, at maximum, $0.5$ for $t \approx 3\times10^{-3} $ and later 
decreases as $k_\mathrm{p}\propto t^{-1/2}$. 
In R1a, which has the lowest chiral feedback parameter,
$k_\mathrm{p}/(7/(4 t)  \, \ln(t/t_\mathrm{log}))^{1/2}$
evolves very similar to the MHD case, R1mhd.
Initially, also $k_\mathrm{p}/(7/(4 t)  \, \ln(t/t_\mathrm{log}))^{1/2}$
in R1 evolves similar as in R1mhd.
But at $t \approx 0.2$, the transition to phase (iii) 
occurs and $k_\mathrm{p}/(7/(4 t)  \, \ln(t/t_\mathrm{log}))^{1/2}$
in R1 evolves similar as in R1b.

%%%%%%%%%%%%%
% Section 4 %
%%%%%%%%%%%%%

\section{Inverse cascade in chiral MHD with turbulence}
\label{sec_highRe}

In this section we explore the transition from laminar to
turbulent flows.
In particular, we are interested in how turbulence modifies the three-phase scenario
of a decaying helical magnetic field in chiral MHD 
that was established in Section~\ref{sec_lowRe}. 
Therefore we run a series of simulations where the 
viscosity and Ohmic resistivity are systematically decreased. 
In the limit of large Reynolds numbers, analytical estimates
can be compared to the results from DNS.

\subsection{Reynolds numbers in DNS of decaying (chiral) MHD turbulence}
\label{subsec_DefRe}

During decaying (chiral) MHD, the magnetic Reynolds number $\Rm$ is a function of time because 
(i) the decaying magnetic field drives a velocity field which changes in time and
(ii) the characteristic wave number on which magnetic forcing occurs corresponds to 
the correlation length of the magnetic field.
The latter increases in time due to the inverse cascade which occurs
when the magnetic field is helical. 
In the following, we approximate the correlation length of the magnetic field
by the scale at which the magnetic energy spectrum reaches its maximum, $\kp$, and
define the time-dependent magnetic Reynolds number as
\begin{eqnarray}
  \Rm(t) = \frac{u_\mathrm{rms}(t)}{\kp(t) \eta}.
  \label{DefReM}
\end{eqnarray}

\begin{figure}
\centering
%/Users/jennifer/Science/Coding/pencil-code/jenny/chiral_fluids/turbulent_decay/postproc_mu50/Rm_t__Rm.py
   \includegraphics[width=0.48\textwidth]{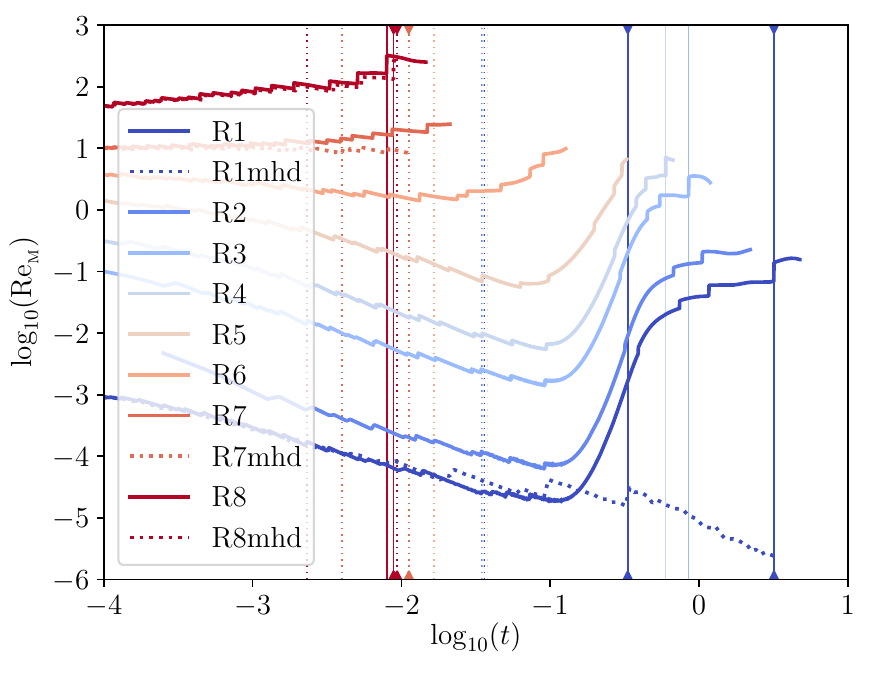} 
\caption{Magnetic Reynolds number $\Rm$ as a function of time.
The colors refer to different simulations of 
classical and chiral MHD simulations as given in the legend;
see Table~\ref{tab_DNSoverview} for details.
In the same colors as the lines showing $\Rm$, we 
indicate as vertical dotted lines the time
at which the minimal magnetic
helicity is reached and as vertical solid lines the time when
the peak of the magnetic energy spectrum $k_\mathrm{p}$ has reached the 
minimal wave number possible in our numerical domain, $k_p=1$, respectively.
Arrows on the vertical lines indicate that these are characteristic times
for classical MHD runs. }
\label{fig_Rm_t__eta}
\end{figure}

The time evolution of $\Rm$ in the majority of simulations 
(all except  R1a, R1b, and R8b) from this study is shown 
in Figure~\ref{fig_Rm_t__eta}.
Especially for the DNS with high diffusion, $\Rm$ changes 
significantly during the simulation time. 
In our reference run for chiral MHD with a vanishing 
velocity field, R1, $\Rm$ decreases from a value of
$\Rm \approx 10^{-3}$ at the beginning to $\Rm \approx10^{-5}$ at 
$t \approx 0.1$, and then increases again, reaching
$\Rm \approx 10^{-1}$ at the final time of the simulation 
$t \approx 4$.
The time dependence in runs where $\Rm$ is larger than
unity in the beginning are less dramatic.
In the most turbulent run, R8, the magnetic Reynolds number
increases only by a factor of approximately $10$.

To distinguish the level of turbulence in different simulations, we
may use the maximum Reynolds number, respectively, which is given as
\begin{eqnarray}
  \Rmmax = \max \left(\frac{u_\mathrm{rms}(t)}{\kp(t) \eta}\right).
\end{eqnarray}
However, $\Rmmax$ is not a useful characteristic of a simulation
because it depends very much on the time at which the simulation is 
stopped. 
A more consistent way of comparing different simulations is by using
Reynolds numbers that are defined at characteristic times during the
evolution. 
In the following, we will use the value of the $\Rm$ at the time 
$t_{k_\mathrm{p}=1}$ at which the peak of the 
magnetic energy spectrum reaches the minimum wave number within
the numerical domain:
\begin{eqnarray}
  \Rmkpkone = \Rm(t=t_{k_\mathrm{p}=1}).
\end{eqnarray}
Additionally, we will consider the Reynolds number at the time 
$t_{\mathrm{min}(\mathcal{H})}$
when the magnetic helicity reaches its minimum:
\begin{eqnarray}
  \RmABmin = \Rm(t=t_{\mathrm{min}(\mathcal{H})}).
\end{eqnarray}
The values of $\Rmmax$, $\Rmkpkone$, and $\RmABmin$ for all
DNS presented in this work
are listed in the last three columns of 
Table~\ref{tab_DNSoverview}.

\begin{figure}
\centering
%/Users/jennifer/Science/Coding/pencil-code/jenny/chiral_fluids/turbulent_decay/postproc_mu50/energy_t__Rm.py
   \includegraphics[width=0.48\textwidth]{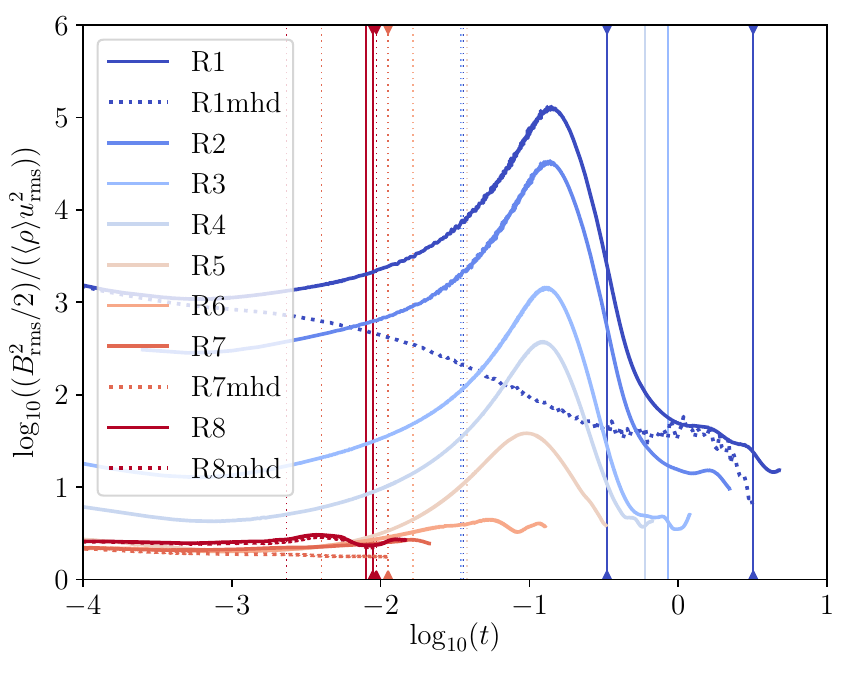} 
\caption{Ratio of magnetic over kinetic energy as a function of time
for the same simulations as presented in Figure~\ref{fig_Rm_t__eta}.
See the caption of Figure~\ref{fig_Rm_t__eta} for a description of
the thin vertical lines.
}
\label{fig_energy_t__eta}
\end{figure}

\begin{figure}
\centering
%/Users/jennifer/Science/Coding/pencil-code/jenny/chiral_fluids/turbulent_decay/postproc_mu50/kpcorr_t__Rm.py
   \includegraphics[width=0.48\textwidth]{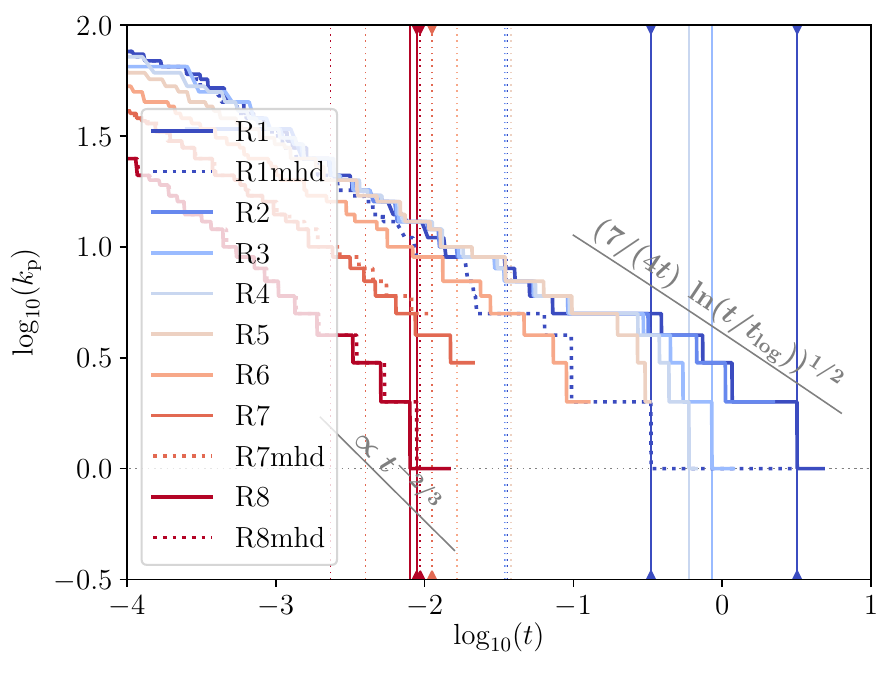} 
\caption{The peak scale of the magnetic energy spectrum
as a function of time for the same simulations as presented 
in Figure~\ref{fig_Rm_t__eta}. 
See the caption of Figure~\ref{fig_Rm_t__eta} for a description of
the thin vertical lines.
}
\label{fig_scaling_t}
\end{figure}

\subsection{Transition from low to high $\Rm$ in DNS of decaying (chiral) MHD turbulence}

\begin{figure}[th!]
  \centering
%
% pencil-code/jenny/chiral_fluids/turbulent_decay/postproc_mu50/
\includegraphics[width=0.48\textwidth]{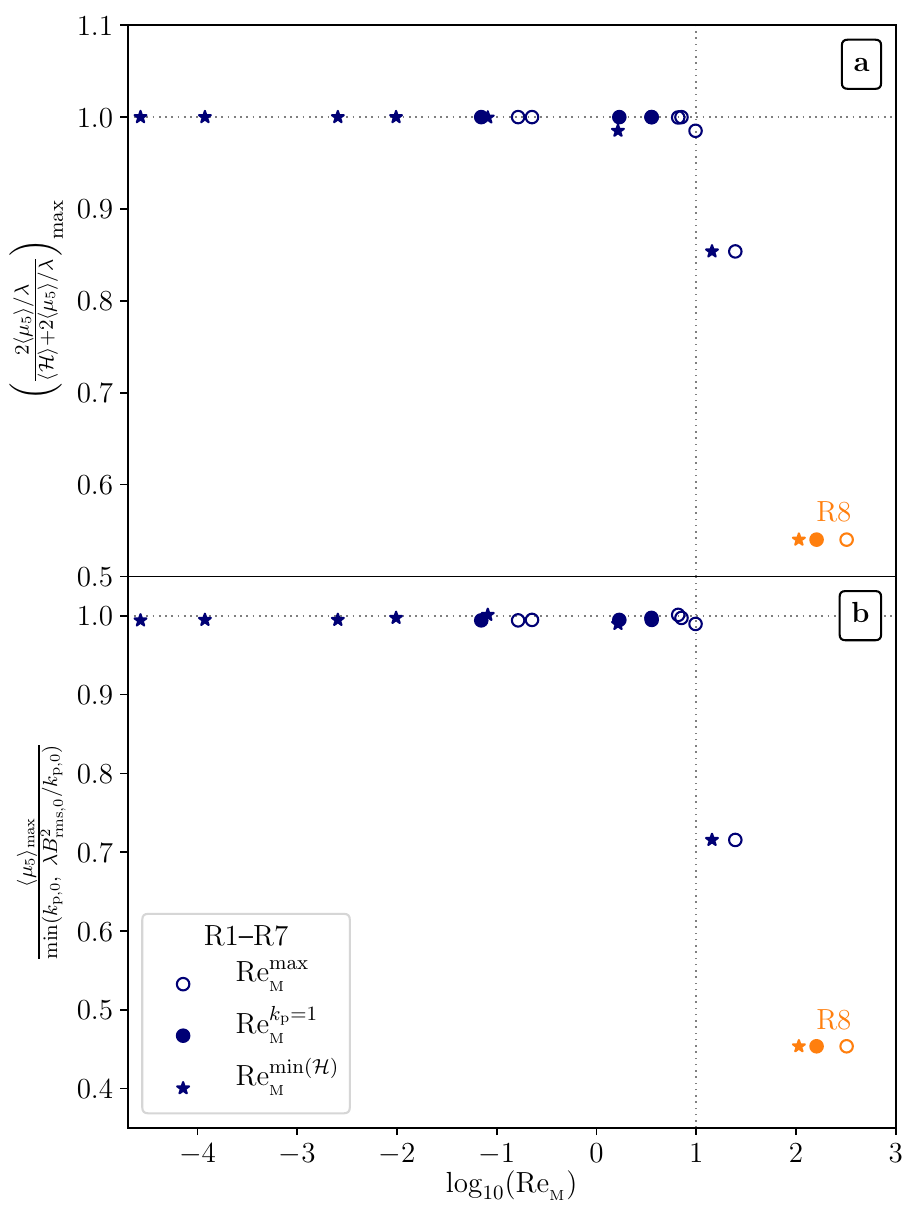}
\caption{Comparing runs R1, R2, R3, R4, R5, and R6, presented by 
blue data points. 
In orange color, run R8 is shown which has the largest $\Rm$.
Note, however, that R8  
has a different value of $\lambda B_0^2/(k_{\mathrm{p},0})$ in comparison to 
R1--R6; see Table~\ref{tab_DNSoverview}.
Key properties of the runs are plotted as a function of the magnetic Reynolds number
$\Rm$ for which three characteristic values are presented:
$\Rmkpkone$ is the magnetic Reynolds number at the time
when the peak of the magnetic energy spectrum reaches the minimum value in
the numerical domain, $\RmABmin$ is
the magnetic Reynolds number at the time when the minimum of 
$\langle\mathcal{H}\rangle$ is reached, and $\Rmmax$
is the maximum Reynolds number during the entire simulation time. 
The latter depends strongly on time when the simulation is stopped and is 
not suitable for a comparison between different runs.  \\
\textit{a)} Maximum value of $2\langle\mu_5\rangle/\lambda$ over the conserved 
quantity $\langle\mathcal{H}\rangle +2 \langle\mu_5\rangle/\lambda$.\\
\textit{b)} Maximum value of $\langle\mu_5\rangle$ generated in the 
simulation over the theoretically predicted value in the kinematic limit 
$\mathrm{min}(k_{\mathrm{p},0}, \lambda B_0^2/k_{\mathrm{p},0})$.
}
\label{fig_Rm1}
\end{figure}

\begin{figure}[th!]
  \centering
%
% pencil-code/jenny/chiral_fluids/turbulent_decay/postproc_mu50/
\includegraphics[width=0.48\textwidth]{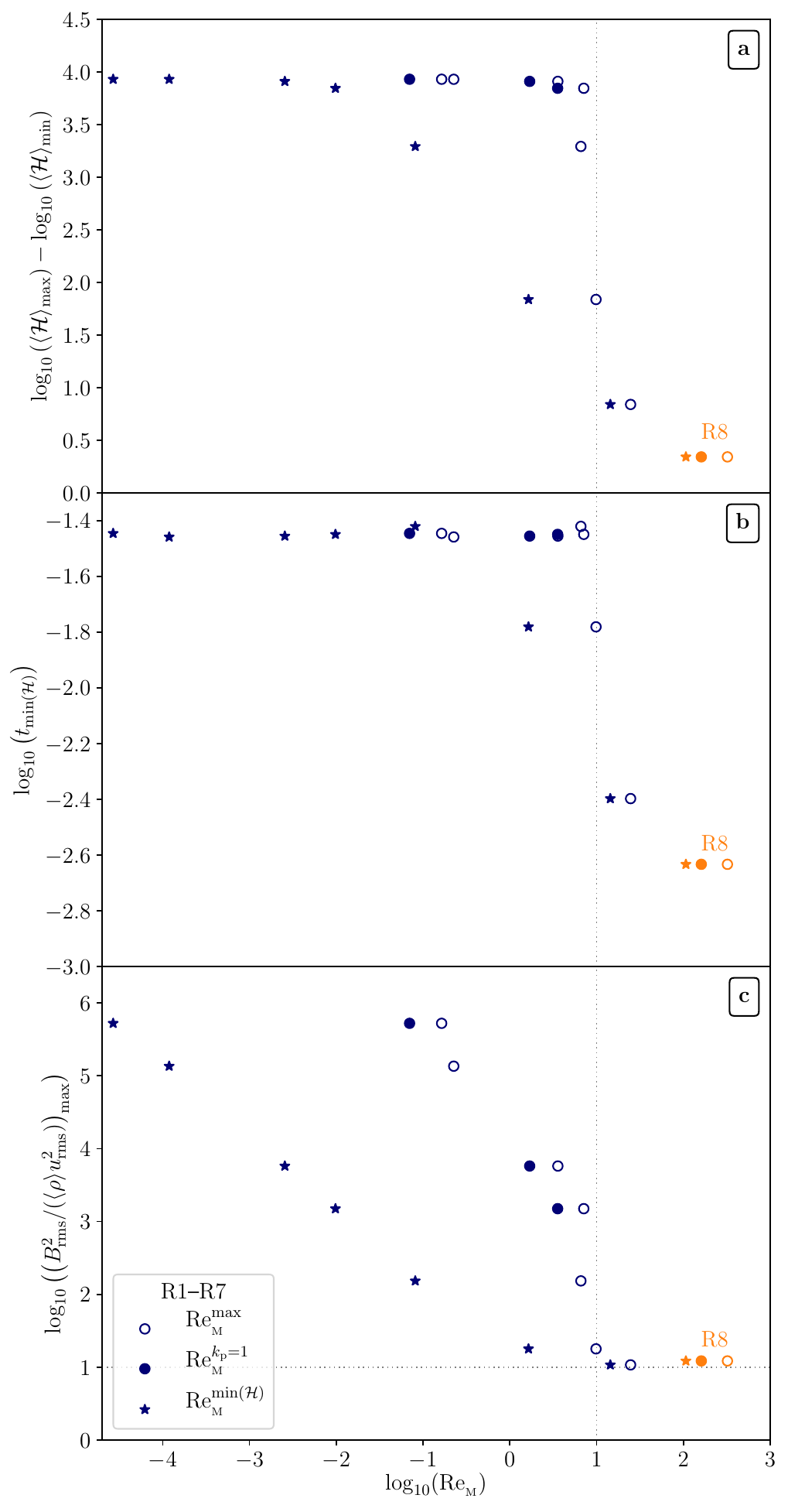}
\caption{Using the same conventions as for Figure~\ref{fig_Rm1} we show:\\
\textit{a)} Difference between the maximum and the 
minimum value of the logarithm of the magnetic helicity  $\langle\mathcal{H} \rangle$.\\
\textit{b)} Time until the minimum of $\langle\mathcal{H}\rangle$
is reached $t_{\mathrm{min}(\bf{A}\cdot\bf{B})}$.\\
\textit{c)} Maximum of the ratio of magnetic over kinetic energy for $t>t_{\mathrm{min}(\bf{A}\cdot\bf{B})}$.
}
\label{fig_Rm2}
\end{figure}

Starting from our reference run of chiral MHD with negligible velocity
field, R1, we systematically decrease the 
values of $\eta$ and $\nu$ in runs R2--R8 in order to explore
the transition to the turbulent regime where the velocity field
is expected to impact significantly the evolution of 
the magnetic field. 
As a characteristic parameter for the degree of 
nonlinearity in the Equations~(\ref{eq_ind})--(\ref{eq_mu5D}),
we list the ratio of the initial magnetic field strength
$B_0$ over $\eta$ for all DNS
in the 7th column of Table~\ref{tab_DNSoverview}.
With $B_0$ being the Alfv\'en velocity within our
unit system, $B_0/\eta$ can be considered 
as the initial Alfv\'enic Reynolds number. 
The time evolution of the $\Rm$ for runs R2--R8
is presented in Figure~\ref{fig_Rm_t__eta}.
The value of $\Rm$ is larger than $1$ in the entire
simulation time of R6, R7, and R8. 
In the latter run, the simulation reaches 
$\Rm \approx 270$. 

When increasing the Reynolds number in the simulations, we observe
two trends. 
First, the maximal ratio of magnetic over kinetic energy density decreases
when $\eta$ and $\nu$ are decreased \footnote{This excludes, of course, the very 
first time step of the simulations where the kinetic energy is zero.}, 
see Figure~\ref{fig_energy_t__eta}.
Second, the scaling of the peak scale of the magnetic energy 
with time
changes from $k_\mathrm{p} \propto (7/(4 t)  \, \mathrm{log}(t/t_\mathrm{log}))^{1/2}$ 
at late times
to $k_\mathrm{p} \propto t^{-2/3}$; see Figure~\ref{fig_scaling_t}.

The changes in the plasma evolution at different Reynolds numbers
is also clearly visible in Figure~\ref{fig_Rm1}.
Different measured characteristics of the simulations
are presented here as a function of $\Rm$. 
According to our discussion in Section~\ref{subsec_DefRe},
for each simulation these parameters are plotted as
a function of $\Rmmax$ (open dots), $\Rmkpkone$ (filled dots), 
and $\RmABmin$ (stars).
In Figure~\ref{fig_Rm1}a, the maximum value of
$2\langle\mu_5\rangle/\lambda$ over the conserved 
total chirality $\langle\mathcal{H}\rangle +2\langle\mu_5\rangle/\lambda$
is presented. 
While $2\langle\mu_5\rangle/\lambda$ is almost 100 percent 
of the total chirality at one time of the plasma 
evolution for $\Rm < 1$, the maximum of the ratio 
$\left(2\langle\mu_5\rangle/\lambda \right)/\left(\langle\mathcal{H}\rangle +2\langle\mu_5\rangle/\lambda\right)$
drops to about $1/2$  once $\Rm$ becomes significantly larger than $1$. 
The mean magnetic helicity, on the other hand, always 
dominates the total chirality at one point in time
for all simulations regardless of their degree of
turbulence as expected from the choice of your initial conditions. 
In simulations with $\Rm>1$, the maximum value of $2\langle\mu_5\rangle/\lambda$ 
is never reached, as is
shown in Figure~\ref{fig_Rm1}b.
The difference between the logarithm of the maximum and the minimum
of $\langle\mathcal{H}\rangle$, see
Figure~\ref{fig_Rm2}a, and also the time needed to reach 
the minimum of 
$\langle\mathcal{H}\rangle$ drops for $\Rm>1$, see 
Figure~\ref{fig_Rm2}b.
Overall, we observe a decrease of the maximal ratio
of magnetic over kinetic energy in our simulations with increasing
magnetic Reynolds number. 
This ratio is plotted in Figure~\ref{fig_Rm2}c and 
decreases continuously with decreasing Ohmic resistivity and
not suddenly at the transition $\Rm\approx1$.

\subsection{Analytical estimates for the limit of large $\Rm$}

Let us estimate analytically the time evolution of $\mu_5$ for  high magnetic
Reynolds numbers.
A high $\Rm$ implies that the first term is more important than
the second term in the right-hand side of Equation~\eqref{eq_ind}.
In the case of the vanishing chiral chemical potential at the initial time, $\mu_{5,0}=0$, as we have seen in the previous section,
the third term never overwhelms the second term in Equation~\eqref{eq_ind},
because $\mu_5(t)$ would decay for $k_{\mathrm{p}}\ll \mu_5$.
Therefore, the evolution of the magnetic field is governed by the interaction to the fluid velocity $\bm{U}$ in the same way as the (nonchiral) classical ideal MHD,
and the magnetic fields undergo the classical inverse cascade.

\begin{figure*}
\centering
  \includegraphics[width=0.96\textwidth]{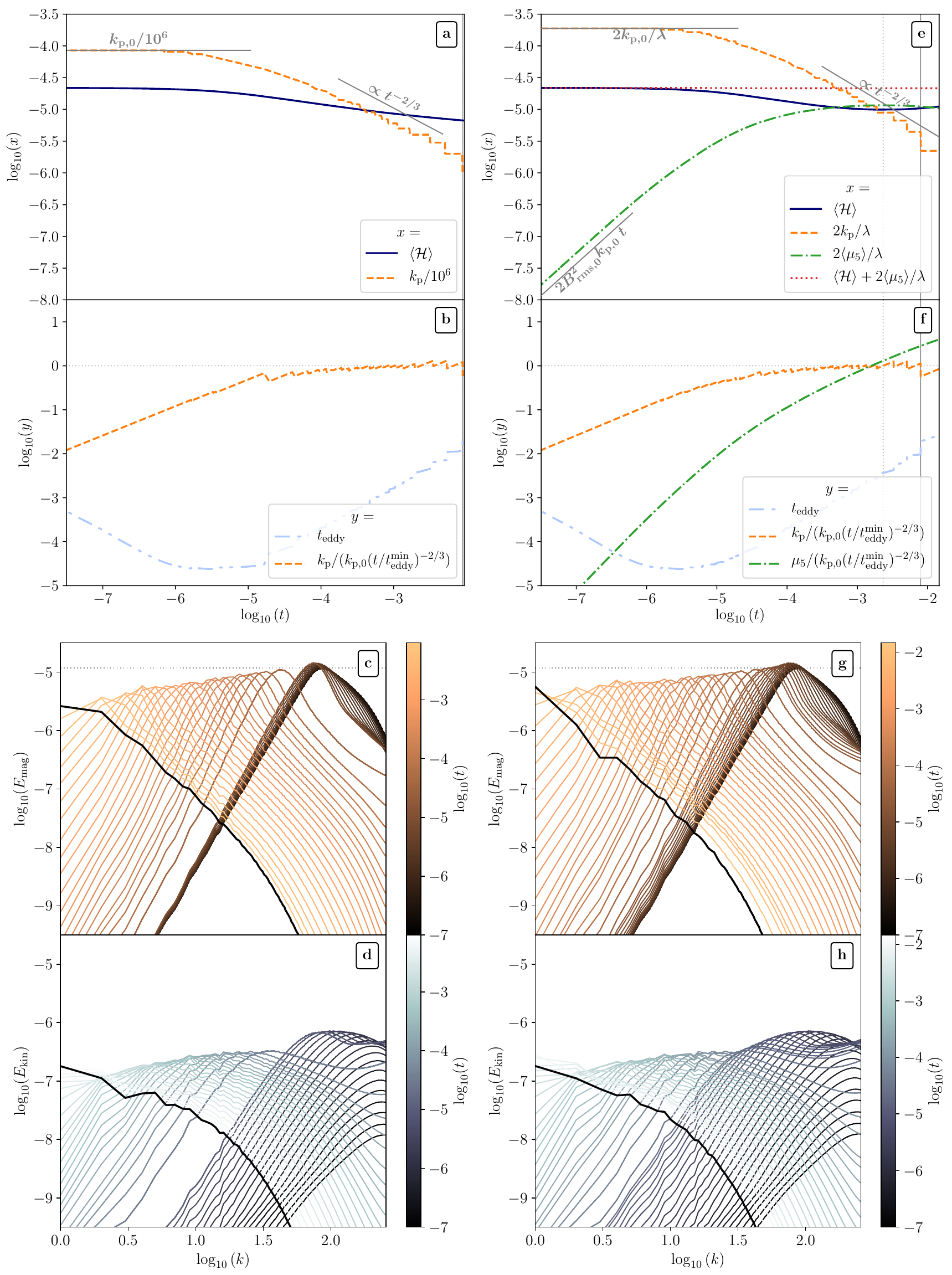}
\caption{Comparing runs with high Reynolds numbers: 
Classical MHD (\textit{left panels}, run R8mhd) 
and chiral MHD (\textit{right panels}, run R8). 
\textit{Top:} Time evolution of the mean magnetic helicity 
$\langle\mathcal{H}\rangle$ and the wave number at which the magnetic energy spectrum
has its maximum $k_\mathrm{p}$. 
For chiral MHD, we also show  $\langle\mu_5\rangle/\lambda$ as well as the conserved quantity 
$\langle\mathcal{H}\rangle +2\langle\mu_5\rangle/\lambda$.
The solid vertical lines indicate the
time when $k_\mathrm{p}$ has reached its
minimum, $k=1$,
and the dotted vertical lines indicate the time
at which $\langle\mathcal{H}\rangle$ is minimal.
\textit{2nd row:} $k_\mathrm{p}$ and $\langle{\mu_5}\rangle$ normalized
by the theoretically expected scaling during self-similar 
evolution (see Section~\ref{sec_analytical}). 
\textit{3rd row:} Evolution of the magnetic energy spectrum. 
The thick dotted black lines show the initial spectra and the 
thick solid black lines show the final spectra of the simulations.
\textit{Bottom:} The same but for the kinetic energy spectrum. 
}
\label{fig_reference_runs_turbulent}
\end{figure*}

As a simple model of the classical inverse cascade~\citep[see, e.g.][]{Durrer:2013pga}, we
consider the following behaviors,
\begin{align}
k_\mathrm{p}(t)&=\left\{\begin{array}{cc}
k_{\mathrm{p},0}\qquad & (t\le t_\mathrm{I}) \\
k_{\mathrm{p},0}(t/t_\mathrm{I})^{-2/3}\quad & (t_\mathrm{I}\le t) \\
\end{array}\right.,
\label{kp IC model}
\\
B(t)&=\left\{\begin{array}{cc}
B_0\qquad & (t\le t_\mathrm{I}) \\
B_0(t/t_\mathrm{I})^{-1/3}\quad & (t_\mathrm{I}\le t) \\
\end{array}\right.,
\label{B IC model}
\end{align}
where $t_\mathrm{I}$ denotes the onset time of the classical inverse cascade.
Inserting this into Equation~\eqref{eq_mu5D} and solving for $\mu_5$, we obtain
\begin{align}
\mu_5(t\le t_\mathrm{I}) & =
k_{\mathrm{p},0}\left[1-e^{-\xi t/t_\mathrm{I}}\right],
\label{eq_mu5_turb_early}\\
\mu_5(t\ge t_\mathrm{I}) & = k_{\mathrm{p},0} \Big[
-3\xi (t/t_\mathrm{I})^{-1/3}
\notag
\\ &
+e^{-3\xi (t/t_\mathrm{I})^{1/3}}      
\Big\{ \left(e^\xi(1+3\xi)-1 \right)e^{2\xi} 
\notag\\
& +9\xi^2 
\left({\rm Ei}(3\xi(t/t_\mathrm{I})^{1/3})-{\rm Ei}(3\xi)\right)\Big\}\Big],
\label{eq_mu5turbtheory2}
\end{align}
where $\xi\equiv \lambda\eta B_0^2 t_\mathrm{I}$ and ${\rm Ei}(x)$ is the exponential integral function.
It is interesting to consider the late time limit of this solution,
\begin{equation}
\mu_5(t\gg \xi^{-3} t_\mathrm{I}) \simeq 
k_{\mathrm{p},0}(t/t_\mathrm{I})^{-2/3},
\label{HiReM mu5}
\end{equation}
which exactly matches $k_{\mathrm{p}}(t)$ in Equation~\eqref{kp IC model}.
Although $\mu_5(t)\approx k_{\mathrm{p}}(t)$ holds at late times 
irrespective of $\Rm$, their time evolution Equations~\eqref{eq_SSE} and \eqref{HiReM mu5} are quite different.
Note that since the sudden change of the behaviors in Equations~\eqref{kp IC model}
and \eqref{B IC model} at $t=t_\mathrm{I}$ are crude approximations, we anticipate
a slight deviation between the analytic estimates of $\mu_5(t)$ and the DNS results there.

\subsection{Simulations of chiral helical MHD turbulence}

% time evolution
In Figure~\ref{fig_reference_runs_turbulent}, the run with lowest diffusion, hence
highest $\Rm$, (R8, left panels) is compared to a classical MHD analogue 
(R8mhd, right panels).
The analysis is exactly the same as in Figure~\ref{fig_reference_runs_laminar},
except for the addition of the kinetic energy spectra in the last row
of Figure~\ref{fig_reference_runs_turbulent}.

Due to the small but finite value of the resistivity in R8mhd,
$\langle\mathcal{H}\rangle$ decays by roughly a factor of three 
over the entire simulation time.
Nevertheless, the magnetic helicity, in combination with
turbulence leads to an efficient inverse cascade in energy
which can be seen in the evolution of the magnetic energy spectrum
in Figure~\ref{fig_reference_runs_turbulent}c. 
The scaling of $\kp$ proportional to $t^{-2/3}$ as expected
for the turbulent inverse cascade of in helical MHD turbulence,
sets in at time $t\approx 10^{-5}$. 
This coincides roughly with the minimum of the eddy turn over time
in the simulation, hence we will use
\begin{eqnarray}
  t_\mathrm{I} \approx 
  t_\mathrm{eddy}^\mathrm{min} \equiv \min \left(\frac{1}{k_\mathrm{p}(t) u_\mathrm{rms}(t)}\right).
\end{eqnarray}
The time evolution of $\kp$ normalized by 
$k_\mathrm{p,0} (t/t_\mathrm{eddy}^\mathrm{min})^{-2/3}$
is presented in Figure~\ref{fig_reference_runs_turbulent}b
The scaling with $t^{-2/3}$ is observed in our DNS for 
times later than approximately $t_\mathrm{eddy}^\mathrm{min}$.
The value of $\kp$ reaches the minimum value of the box 
after $t \lesssim 0.2$.

The time evolution and energy spectra of the chiral MHD run 
with highest magnetic Reynolds number, R8, that are presented in the right
panels of Figure~\ref{fig_reference_runs_turbulent} are very similar 
to the ones in the classical MHD run R8mhd. 
Up $t\approx 10^{-3}$, $k_\mathrm{p}$,
$\langle\mathcal{H}\rangle$, and $t_\mathrm{eddy}$
evolve identically in turbulent MHD and turbulent chiral MHD. 
However, in R8 a $\langle\mu_5 \rangle$ is generated and 
restores a small amount of 
$\langle\mathcal{H}\rangle$. 
The energy spectra in R8mhd and R8 are 
indistinguishable \footnote{Note, that small gaps in time seem
to occur in the energy spectra. This is an artefact from 
the code that calculates energy spectra at a manually fixed 
time interval. The simulations were restarted several time and,
since they cover many orders of magnitude in time, the frequency
of writing spectra is reduced at every restart to save 
computing power. The gap appears 
after a restart when the sampling time step was reduced slightly too 
much to homogeneously fill the logarithmically spaced array of 
example spectra.}.
As expected from Equation~(\ref{eq_mu5_turb_early}),
$\langle\mu_5 \rangle$ increases linearly in time in the beginning.
However the scaling proportional to $t^{-2/3}$
as expected for late times according to Equation~(\ref{HiReM mu5})
is not observed in R8.
This is caused by the fact that the peak of the magnetic energy spectrum 
has moved to the minimum wave number, $k_\mathrm{p}=1$, before 
the scaling of $\langle\mu_5 \rangle$ could converge to
the one of $k_\mathrm{p}$.

\begin{figure}
\centering
%/Users/jennifer/Science/Coding/pencil-code/jenny/chiral_fluids/turbulent_decay/postproc_mu50/energy_t__lambda.py
   \includegraphics[width=0.48\textwidth]{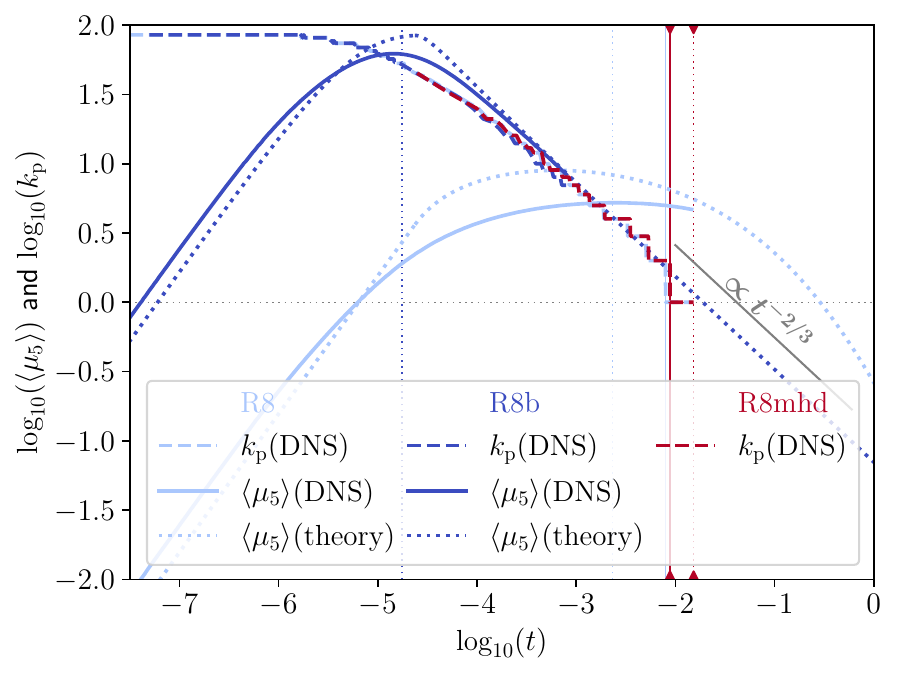} 
\caption{The mean chiral chemical potential, $\langle \mu_5 \rangle$, as
a function of time for runs
R8 and R8b which differ only in their value of 
$\lambda$ (see Table~\ref{tab_DNSoverview}).
Additionally, the evolution of 
$k_\mathrm{p}$
is shown as dashed lines.
For direct comparison also
$k_\mathrm{p}$
from the
corresponding classical MHD run, R8mhd, is presented.
The solid vertical lines indicate the
time when
$k_\mathrm{p}$
has reached the 
minimum wave number possible in the numerical domain, $k=1$,
and the dotted vertical lines indicates the time
at which the minimum of magnetic
helicity is reached.
The horizontal gray dotted line indicates the threshold
for $\mu_5$ above which a dynamo instability occurs
in the numerical box.
}
\label{fig_mu5m_t__lambda_highRm}
\end{figure}

To test the late time scaling of $\langle\mu_5 \rangle \propto t^{-2/3}$
we have repeated run R8 with a larger value of $\lambda$.
For larger $\lambda$, the condition $t\gg \xi^{-3} t_\mathrm{I}$ is 
fulfilled while the 
inverse cascade still proceeds within the numerical domain.
Run R8b has a value of $\lambda$ that is $10^2$ times larger
than the one in R8. 
We compare these two runs with the MHD analog, R8mhd, in 
Figure~\ref{fig_mu5m_t__lambda_highRm}. 
The time evolution of $k_\mathrm{p}$ in all three runs is
almost identical, reaching a scaling of 
$k_\mathrm{p} \propto t^{-2/3}$ at $t\gtrsim 10^{-5}$.
Except for the time around the onset of the inverse
cascade $t_\mathrm{I}$, the time evolution of 
$\langle\mu_5 \rangle$ measured in DNS (solid lines)
agrees very well with the theoretically predicted curves (dotted lines). 
However, the value in DNS is approximately larger by a factor
of $1.5$ compared to the result from 
Equation~(\ref{eq_mu5_turb_early})
at early times.
This behavior might be corrected when including 
the exact shape of the initial magnetic energy spectrum.
The evolution of $\langle\mu_5 \rangle$ in R8b
is very well
described by Equation~(\ref{eq_mu5turbtheory2}) for
$t\gtrsim 10^{-4}$.

%%%%%%%%%%%%%
% Section 6 %
%%%%%%%%%%%%%

\section{Conclusion}
\label{sec_conclusion}

In this study we have explored the evolution of a decaying fully
helical magnetic field in a high-energy plasma in which the 
chiral magnetic effect can occur. 
The chiral magnetic effect is a macroscopic quantum effect that
describes the emergence of an electric current in the presence
of a chiral asymmetry, e.g.\ a difference between the chemical 
potential of left- and right-handed fermions, $\mu_5$, and a magnetic
field. 
The dynamics of such a plasma is determined by the system
of equations of chiral MHD.
We have investigated how a magnetic field decays in chiral
MHD and how a chiral asymmetry develops, from vanishing initial 
$\mu_5$.

When the velocity field plays no dynamically important role,
we have identified a three phase evolution:
(i) a mean chiral chemical potential $\langle \mu_5\rangle$ is produced at 
the expense of the mean magnetic helicity $\meanAB$,
(ii) once $\langle \mu_5\rangle$ exceeds the inverse
correlation length of the helical magnetic field $k_\mathrm{p}$ 
a chiral dynamo reestablishes $\meanAB$,
and
(iii) a self-similar inverse cascade where 
$|\mu_{5}(t)| \approx   k_\mathrm{p}(t) \approx \left[(3 + n)/(4 \eta t)~\ln\left(t/t_\mathrm{log}\right)\right]^{1/2}$,
where $n$ is the slope of the initial magnetic helicity spectrum.
A similar scenario has been reported by \citet{HironoEtAl2015}.
However, they did not include the logarithmic correction in phase (iii)
that agrees well with our DNS.

Our simulations, performed with the \textsc{Pencil Code} that
has explicit viscosity and Ohmic resistivity, allow us to
systematically explore a decaying magnetic field in
chiral MHD and, in particular, to probe
the transition from low to high 
magnetic Reynolds numbers.
We find that the magnetic energy spectrum evolves more and more
similar in chiral MHD and classical MHD when $\Rm$ is increased
by lowering the dissipation in DNS.
For $\Rm \gg 1$, we observe an inverse cascade of the magnetic
field, where the peak of the magnetic energy spectrum
$k_\mathrm{p} \propto t^{-2/3}$, regardless of the existence of 
an additional degree of freedom in from of a $\mu_5$.
In the simulations of chiral MHD, a $\mu_5$ is initially
generated linearly in time. 
For $t\gg \xi^{-3} t_\mathrm{I}$ with $\xi\equiv \lambda\eta B_0^2 
t_\mathrm{I}$ we find a scaling of
$\mu_5 \simeq k_{\mathrm{p},0}(t/t_\mathrm{I})^{-2/3}$, 
i.e., like for low Reynolds number, the evolution of $\mu_5$ follows 
the one of $k_\mathrm{p}$, $k_\mathrm{p} \simeq \mu_5$.

We have shown that the chiral magnetic effect not only modifies the 
inverse cascade of magnetic fields but it also leads to the generation 
of a chiral anomaly. 
Such an anomaly may manifest itself in the neutrino density in the 
late Universe and it might be relevant, e.g., for the the number of 
effective degrees of freedom, $N_{\rm eff}$, at recombination. 
A study of this possibility is referred to future work.
Our results may also have other important implications for the
evolution of  relativistic plasmas in the early Universe
and protoneutron stars.

%%%%%%%%%%%%%%%%%%%
% Acknowledgments %
%%%%%%%%%%%%%%%%%%%

\begin{acknowledgments}

The authors would like to acknowledge the Mainz Institute for Theoretical Physics (MITP) of the
DFG Cluster of Excellence PRISMA$^+$ (Project ID 39083149), for enabling us to complete a significant 
portion of this work.
JS acknowledges the funding from the 
Swiss National Science Foundation under Grant No.\ 185863,
European Union's Horizon 2020 research and
innovation program under the Marie Sk{\l}odowska-Curie grant
No.\ 665667, and
the support by the National Science Foundation under Grant No.\ NSF PHY-1748958.
The work of TF was supported by JSPS KAKENHI No. 17J09103 and No.\ 18K13537.
RD is supported with the Swiss National Science Foundation under Grant No. 200020\underline{~}182044.

The simulations presented in this work were performed on resources at Chalmers Centre 
for Computational Science and Engineering (C3SE) provided by the Swedish National 
Infrastructure for Computing (SNIC) as well as on the Baobab cluster at the University of Geneva.

\end{acknowledgments}

%%%%%%%%%%%%
% Appendix %
%%%%%%%%%%%%

\appendix

\section{Derivation of the self-similar evolution in the limit of a vanishing velocity field}
\label{sec_SSEderivation}

Here we shall derive the self-similar solution of $\mu_5(t)$ and $k_\mathrm{p}(t)$ in the phase (iii), Equation~\eqref{eq_SSE}.
Ignoring the velocity field in Equation~\eqref{eq_ind}, 
the equation of motion for magnetic field reads 
\begin{equation}
\partial_t \bm B=\eta\bm 
\nabla\times\left(- \bm \nabla\times \bm B+\mu_5 \bm B \right).
\label{B EoM Appendix}
\end{equation}
The Fourier transformation of the magnetic field is written as
\begin{equation}
\bm B(t,\bm x)=\sum_{\lambda=\pm}
\int \frac{{\rm d}^3 k}{(2\pi)^3} 
 e^{i \bm{k \cdot x}} 
 \bm{e}_\lambda(\hat{\bm k})
 B_\lambda(t,k),
\end{equation}
where $\lambda=\pm$ is the label of the circular polarization 
and $\bm{e}_\lambda(\hat{\bm k})$ is the circular polarization vector that satisfies $i \bm{k} \times \bm{e}_\pm(\hat{\bm k}) = \pm k \bm{e}_\pm(\hat{\bm k})$.
Using the relation between the magnetic helicity and $B_\pm$ in Fourier space,
$\mathcal{H}_k\equiv\left(B_+^2-B_-^2 \right)/k$, one can recast Equation~\eqref{B EoM Appendix} into the equation of motion for the magnetic helicity, 
\begin{equation}
\partial_t \mathcal{H}_k+2\eta k^2 \mathcal{H}_k
-4\eta\mu_5 \, \rho_B(k)=0,
\end{equation}
where the spatial fluctuations of $\mu_5$ are neglected (i.e.\ $\mu_5\rightarrow \langle\mu_5\rangle$) and $\rho_B(k)\equiv (B_+^2+B_-^2)/2$ is the magnetic energy density in Fourier space.
When the magnetic field is maximally helical, $B_\pm^2\gg B_\mp^2$, $\rho_B(k) 
\simeq \sigma_{\mathcal{H}}\, k\mathcal{H}_k/2$, where $\sigma_{\mathcal{H}}\equiv {\rm sign}[\mathcal{H}_k]$ ensures $\rho_B$ is always positive.
In this case, the above equation is reduced into
\begin{equation}
\partial_t \mathcal{H}_k+2k\eta( k -|\mu_5|) \mathcal{H}_k
= 0,
\quad
{\rm (maximal\, helical)}
\label{max hel eq}
\end{equation}
where we used $\sigma_{\mathcal{H}}\mu_5=|\mu_5|$ because the magnetic helicity and $\mu_5$ have the same sign, if $\mu_5$ is produced by the magnetic helicity as we assume.
This equation has the formal analytic solution
\begin{equation}
\mathcal{H}_k(t)=\mathcal{H}_k(t_0)\exp\left[
2k\eta\left(-k (t-t_0) +\theta(t)\right)
\right],
\end{equation}
with $\theta(t)\equiv \int^{t}_{t_0} \dd t' |\mu_5(t')|.$
In the phase (iii) the magnetic helicity dominates the conserved quantity
and hence $\mathcal{H} (\gg 2\mu_5/\lambda)$ is independently conserved 
\begin{equation}
\int \dd k\, k^2 \mathcal{H}_k(t_0)\exp\left[
2k\eta\left(-k (t-t_0) +\theta(t)\right)
\right]={\rm const}.
\end{equation}
Taking the time derivative of this equation and 
dropping time dependent but nonvanishing factors, one finds
\begin{equation}
\int^{k_*}_0 \dd k\, k^{3+n}
\left[ |\mu_5(t)|-k  \right]
e^{-2\eta (t-t_0)\left(k-\frac{\theta(t)}{2(t-t_0)}\right)^2}=0.
\end{equation}
Here we assume a power-law helicity slope with an UV-cutoff at $k_*$
\begin{equation}
\mathcal{H}_k(t_0)=\mathcal{H}_{k_*}(t_0) \left(\frac{k}{k_*}\right)^n
\Theta(k_*-k),
\end{equation}
where $\Theta(x)$ is the Heaviside function.
Changing the dummy variable from $k$ into $p\equiv \sqrt{2\eta (t-t_0)}k$, we obtain
\begin{equation}
|\mu_5(t)|=\frac{1}{\sqrt{2\eta (t-t_0)}}\,
\frac{\int^{p_*}_0\dd p\, p^{4+n} e^{-\left(p-\sqrt{\frac{\eta}{2(t-t_0)}}\,\theta(t)\right)^2}}
{\int^{p_*}_0\dd p\, p^{3+n} e^{-\left(p-\sqrt{\frac{\eta}{2(t-t_0)}}\,\theta(t)\right)^2}},
\label{mu5 exact}
\end{equation}
with $p_*\equiv \sqrt{2\eta (t-t_0)}k_*$.

To simplify this expression, we make an additional approximation. 
For this, we restrict ourselves into a late time regime, $t \gg t_0$. 
Then the upper limit of the integrals $p_* \propto t^{1/2}$ can be approximated by $\infty$,
and $\sqrt{\eta/2t}\,\theta$ is also considered as large,
as we will confirm a posteriori.
The integrals are computed for $X\equiv\sqrt{\eta/(2(t-t_0))}\,\theta(t)\gg 1$ as
\begin{equation}
\frac{\int^{\infty}_0\dd p\, p^{4+n} \exp\left[-\left(p-X\right)^2\right]}{\int^{\infty}_0\dd p\, p^{3+n} \exp\left[-\left(p-X\right)^2\right]}
= X +\frac{3+n}{X}+\mathcal{O}(X^{-3}).
\end{equation}
Therefore Equation~\eqref{mu5 exact} is simplified to
\begin{equation}\label{e:dtheta}
   \theta'(t)= |\mu_5(t)| \simeq \frac{\theta(t)}{2t} + 
\frac{3+n}{2\eta \theta(t)}.
\end{equation}
Note that in order for $\mathcal{H}$ not to develop an infrared singularity we must require $n>-3$ so that both terms in (\ref{e:dtheta}) are always positive.
The solution of this differential equation is %
\begin{align}
\theta(t)&\simeq \sqrt{\frac{t}{\eta}}\sqrt{\mathcal{C}+(3+n)\ln\left(\frac{t}{t_\mathcal{C}}\right)},
\\
\mu_5(t)&\simeq \frac{1}{2\sqrt{\eta t}}\frac{3+n+\mathcal{C}+(3+n)\ln(t/t_\mathcal{C})}{\sqrt{\mathcal{C}+(3+n)\ln(t/t_\mathcal{C})}},
\end{align}
where $\mathcal{C}$ is an integration constant and $t_\mathcal{C}$ is degenerate with $\mathcal{C}$.
The approximation used above, $X= \sqrt{\eta/2t}\,\theta \gg 1$, is valid for a sufficiently late time,
\begin{equation}
\sqrt{\frac{\eta}{t}}\theta(t)\simeq
\sqrt{\mathcal{C}+(3+n) \ln\left(\frac{t}{t_\mathcal{C}}\right)} \gg 1. 
\end{equation}
This also allows us to further simplify $\mu_5$ as
\begin{equation}
\mu_5\simeq 
\left[\frac{3+n}{4\eta t}\ln\left(\frac{t}{t_{\rm log}}\right)\right]^{\frac{1}{2}},
\label{mu5 latetime result}
\end{equation}
where the integration constant is rewritten as $\mathcal{C}=(3+n)\ln(t_\mathcal{C}/t_{\rm log})$, assuming $n\neq -3$.
Note that this logarithmic correction which slightly slows down the decay of $\mu_5$
becomes more significant as $n$ increases.
It implies that also the inverse cascade (i.e. the transportation to larger scales) of the peak scale $k_p$  is slowed-down.
This is because it takes more time for a large-scale helicity modes to grow large enough to ensure the conservation law, when the initial helicity has a bluer spectrum, i.e. more power on smaller scales. 
If the initial helicity is scale invariant, $n=-3$, the logarithmic correction vanishes.

%%%%%%%%%%%%
% references %
%%%%%%%%%%%%

\bibliography{references}

\end{document}